\documentclass[aip,jcp]{revtex4-2}

\usepackage{makecell}
\usepackage{xcolor}
\usepackage{float}
\usepackage{graphicx}
\usepackage{tikz}
\usepackage{hyperref}
\usepackage{algorithm2e}
\usepackage{algorithmic}
\usepackage{longtable}
\usepackage{amsmath,amssymb}

\definecolor{fitzred}{RGB}{139,0,0}
\definecolor{fitzgray}{RGB}{70,70,70}
\DeclareMathOperator{\E}{\mathbb{E}}

\newcommand{\degree}{$^{\circ}$}
\newcommand{\rsq}{$\textit{R}^2=$\space}

\draft % marks overfull lines with a black rule on the right

\begin{document}

%\preprint{}

\title{Enhancing NEMD with automatic shear rate sampling to model viscosity and correction of systematic errors in modelling density: Application to linear and light branched alkanes} %Title of paper



\author{Pavao Santak}
\email[Corresponding author:]{ ps727@cam.ac.uk}
\author{Gareth Conduit}
%\thanks{}
\affiliation{Theory of Condensed Matter, Department of Physics, University of Cambridge, J.J.Thomson Avenue, Cambridge, CB3 0HE, United Kingdom}


\date{\today}

\begin{abstract}
We perform molecular dynamics simulations to model density as a function of temperature for 74 alkanes with 5 to 10 carbon atoms and non-equilibrium molecular dynamics simulations in the NVT ensemble to model kinematic viscosity of 10 linear alkanes as a function of molecular weight, pressure, and temperature. To model density, we perform simulations in the NPT ensemble before applying correction factors to exploit the systematic error in the SciPCFF force field, and compare results to experimental values, obtaining an average absolute deviation of 3.4 $\frac{\rm g}{\rm l}$ at 25\degree C and of 7.2 $\frac{\rm g}{\rm l}$ at 100\degree C. We develop a sampling algorithm that automatically selects good shear rates at which to perform viscosity simulations in the NVT ensemble and use Carreau model with weighted least squares regression to extrapolate Newtonian viscosity. Viscosity simulations are performed at experimental densities and show an excellent agreement with experimental viscosities, with an average percent deviation of -1\% and an average absolute percent deviation of 5\%. Future plans to study and apply the sampling algorithm are outlined. 
\end{abstract}

\maketitle %\maketitle must follow title, authors, abstract


\section{Introduction}
%\label{}
Alkanes are of great interest to both the academic community and a large number of scientists and engineers using them in industry. Their chemical simplicity makes them an ideal testing ground for applications of novel computational methods in studying physical properties of complex fluids. In industry, understanding alkanes and their properties is essential to produce superior oil and gas products. 

One of the most important properties of alkanes is kinematic viscosity, which is a measure of their flow properties. However, the viscosity of pure alkanes is still poorly understood. While many viscosity measurements of mixtures are made in industrial laboratories on a daily basis, the viscosity of only about 20 pure alkanes has been published in the academic literature, and difficulties in separation of different isomers beyond dodecane prevent engineers and scientists from making measurements of viscosity of large alkanes. Consequently, several theoretical and computational methods have been developed to investigate alkanes' viscosity variation with molecular structure, temperature, and external pressure. For example, De La Porte and Kossack modelled viscosity of long chain n-alkanes with a model motivated by the free volume theory \cite{ViscosityComparison}; Riesco and Vesovic used a hard sphere model to predict viscosity of similar systems\cite{HardSphere}, and Novak modelled viscosity of alkanes with a corresponding states model \cite{Lawrence}. Modern statistical methods have also been used to model viscosity of alkanes. Santak and Conduit modelled kinematic viscosity of n-alkanes with a neural network that can make predictions on sparse datasets \cite{me}; Suzuki et al. utilized fully connected neural networks to model viscosity as a function of temperature of various organic compounds \cite{ANNSuzuki}, while Hosseini et al. used a neural networks and a hard sphere model to model similar systems\cite{ANNHoss}. 

Equilibrium molecular dynamics (EMD), frequently applied to model viscosity of light alkanes, is another popular computational method. Cui et al. modelled viscosity of hexadecane, tetracosane, and decane \cite{4}, and compared molecular and atomic formalisms for EMD simulations of decane \cite{4}; Singh, Payal et al. modelled viscosity of hexadecane with several force fields \cite{5}; Zhange and Ely modelled viscosity of alkane systems and alcohols \cite{12}, while Kondratyuk modelled viscosity of triacontane \cite{25}. Furthermore,  Kioupis and Maginn modelled viscosity of a hexane/hexadecane mixture \cite{Kioupis1}, and determined the viscosity number in addition to investigating viscosity variation with pressure of three distinct poly-$\alpha$-olefins \cite{Kioupis2} \space \cite{Kioupis3}, while Mundy et al. predicted viscosity of n-decane, n-hexadecane, 6-pentylundecane, 7,8-dimethyltetradecane, 2,2,4,4,6,8-heptamethylnonane, n-triacontane and squalane \cite{Mundy2} and determined pressure-viscosity coefficient of decane \cite{Mundy3}.

Nevertheless, none of the semi-analytical methods, the modern statistical methods and EMD have been certified to reliably model the viscosity of all alkanes. Semi-analytical methods do not possess enough sufficient predictive power to be judiciously extrapolated to alkanes outside of the training set, which usually comprises a limited set of light alkanes. Modern statistical methods possess greater extrapolative power than their semi-analytical counterparts, yet their utility is still limited by the lack of experimental data. EMD can in principle be used for all alkanes, but because of slow relaxation of the stress-stress autocorrelation function \cite{3} \space \cite{4} for larger molecules \cite{2} it is recommended to primarily use it in modelling low viscosity molecules \cite{2}.

Another physics based simulation method that has gained momentum in the past several decades is the non-equilibrium molecular dynamics (NEMD) \cite{19}, in which shear is applied to a molecular system, usually at fixed temperature and volume. A molecular dynamics simulation is performed at several shear rates, and the shear rate profile of the kinematic viscosity is then extrapolated to Newtonian viscosity. In addition to applying EMD, Kioupis and Maginn also used NEMD to model viscosity of hexane/hexadecane binary mixture \cite{Kioupis1}, and of three poly-$\alpha$-olefins \cite{Kioupis2} \space \cite{Kioupis3}, while Mundy et al. utilised NEMD to study viscosity of decane \cite{Mundy1} and several large branched alkanes \cite{Mundy2}. Cui et al. used NEMD to model viscosity of decane at 25$^{\circ}$C, hexadecane at 27$^{\circ}$C and 50$^{\circ}$C, tetracosane and 10-hexylnonadecane at 60$^{\circ}$C, and squalane at 39$^{\circ}$C and 99$^{\circ}$C \cite{1} \cite{4}; McCabe, Pan, and Evans modelled viscosity of decane \cite{3} \cite{9}; Liu et al. modelled viscosity of squalane and 1-decene-trimer \cite{8}; Cho, Jeong, and Buig modelled viscosity of polymer melts \cite{15}; Yang, Pakkanen, and Rowley determined viscosity index of various lubricant size molecules \cite{16}, as well as of several small alkane mixtures \cite{18}; Liu et al. determined a pressure viscosity coefficient of a 1-decene trimer \cite{22}; Allen and Rowley compared different force fields to model viscosity of small alkanes \cite{21}, while Khare, de Pablo and Yethiraj modelled viscosity of hexadecane, docosane, octacosane and 5,12-dipropyl-hexadecane \cite{23} and J.D Moore,  S.T Cui, H.D Cochran and P.T Cummings modelled viscosity of C100 \cite{C100} . 

However, despite its past success in modelling viscosity of some alkanes, the contemporary NEMD approach still suffers from three pitfalls. Firstly, any viscosity simulation result carries a systematic error from the force field that determines the motion of atoms and molecules. Secondly, to perform NEMD simulations at accurate external conditions, the density of the alkane of interest needs to be either experimentally known or accurately modelled with molecular dynamics. Despite possessing more experimental data for density than for viscosity, the density of most alkanes is experimentally unknown, and while molecular dynamics simulation results are frequently used to replace experimental values, they need to be in close agreement with true values to be confidently applies as initial conditions in NVT simulations; otherwise, simulations are performed at a wrong external pressure and viscosity simulation results will carry a large systematic error due to viscosity's pressure dependence. Finally, the reliability of viscosity simulations decreases, while uncertainty in viscosity simulation results increases with decreasing shear rate, making direct identification of Newtonian viscosity difficult, with its accurate extrapolation dependent on performing the simulations at appropriate shear rates. Currently, no computational method is capable of systematically and automatically selecting good simulation shear rates for any alkane at arbitrary external conditions.

In this manuscript, we present two computational techniques that enhance the current NEMD method. Firstly, we split alkanes into several groups and apply correction factors to each simulation result to correct errors in density predictions. Secondly, we develop a sampling algorithm that automatically samples good shear rates and apply the weighted least squares regression to extrapolate Newtonian viscosity. In \autoref{2}, we present simulation methodology and in \autoref{3}, we model liquid density of small linear, single-branched and double branched alkanes, and kinematic viscosity of n-alkanes as a function of molecular weight, temperature, and pressure. To model viscosity, we perform simulations at experimental densities values to directly assess the performance of the sampling algorithm. Finally, in \autoref{4}, we succinctly summarize our work and outline the plans for future development and applications of methods presented in the foregoing sections. 

Experimental data for density and viscosity was obtained from the TRC Thermodynamic Tables \cite{TRC}, with additional viscosity data collected from several research papers \cite{doi:10.1021/je800417q} \space \cite{Caudwell2004} \space \cite{TridecanePressure}. Input files for simulations are prepared in the MAPS platform (\url{https://www.scienomics.com/}) and performed in LAMMPS \cite{LAMMPS}. To compare simulations to experiments, we use the average absolute deviation ($\left|\Delta\right|$) for density and percent error ($\Delta_{\%}$) together with the absolute percent error ($|\Delta_{\%}|$) for viscosity, chosen for their interpretability and widespread use in the literature.

\section{Simulation details}\label{2}
Molecular dynamics is a computational simulation technique in which empirically parameterized force fields determine the interactions and govern the equations of motion for atoms and molecules. Due to relative simplicity in performing simulations for diverse physical systems rather than experimentally creating them, molecular dynamics can provide predictive capability and novel insights into the properties of physical systems that have not yet been experimentally produced. 

In this section, we describe simulation techniques implemented to model liquid density and kinematic viscosity of alkanes. In \autoref{2.1}, we outline the density simulation procedure, and in \autoref{2.2}, we describe data blocking, which enables us to accurately determine uncertainty in simulated physical quantities. Then, in \autoref{2.3}, we introduce simulation details to calculate viscosity, and in \autoref{2.4} we describe a sampling algorithm that automatically identifies and samples the shear rates at which to perform the simulations.

\subsection{Molecular dynamics density simulations}\label{2.1}
To model density, molecular dynamics simulations are performed in the NPT ensemble to simulate real experimental conditions. Simulation input files are prepared by building a molecule in MAPS, optimizing its geometry and applying a SciPCFF force field, which is a Scienomics (\url{https://www.scienomics.com/}) implementation of the PCFF\cite{PCFF} force field with COMPASS\cite{COMPASS} parameters. Next, we build a cubical unit cell with a side length of 40\AA \space and density of 800$\frac{g}{l}$ at the simulation temperature before applying periodic boundary conditions. The cell geometry is then optimized by minimizing its energy for 500 time steps with a conjugate gradient to yield the best simulation initial conditions.

We apply a 12\AA \space cutoff without smoothing to the force field and tail corrections to the Van der Waals interactions. For the Coulomb interaction, we use a particle mesh, a precision of 0.0001 and a dielectric constant of 1, but do not apply a cutoff to it. To keep the system at a constant temperature and pressure, we implement a Nose-Hoover thermostat/barostat with a 10fs temperature damping and a 350fs pressure damping. We perform simulations for 1ns with a time step of 1fs, and take a volume measurement taken every 1000 time steps. Equations of motion are integrated with the velocity Verlet algorithm. Simulations could be sped up at the potential expense of lower simulation accuracy by using a multi-step algorithm or applying constraints, but here we focus on enhancing the existing methods with improved sampling of shear rates and defer further optimization to future work.

\begin{figure}[H]
	\centering
	\includegraphics[width=0.55\linewidth]{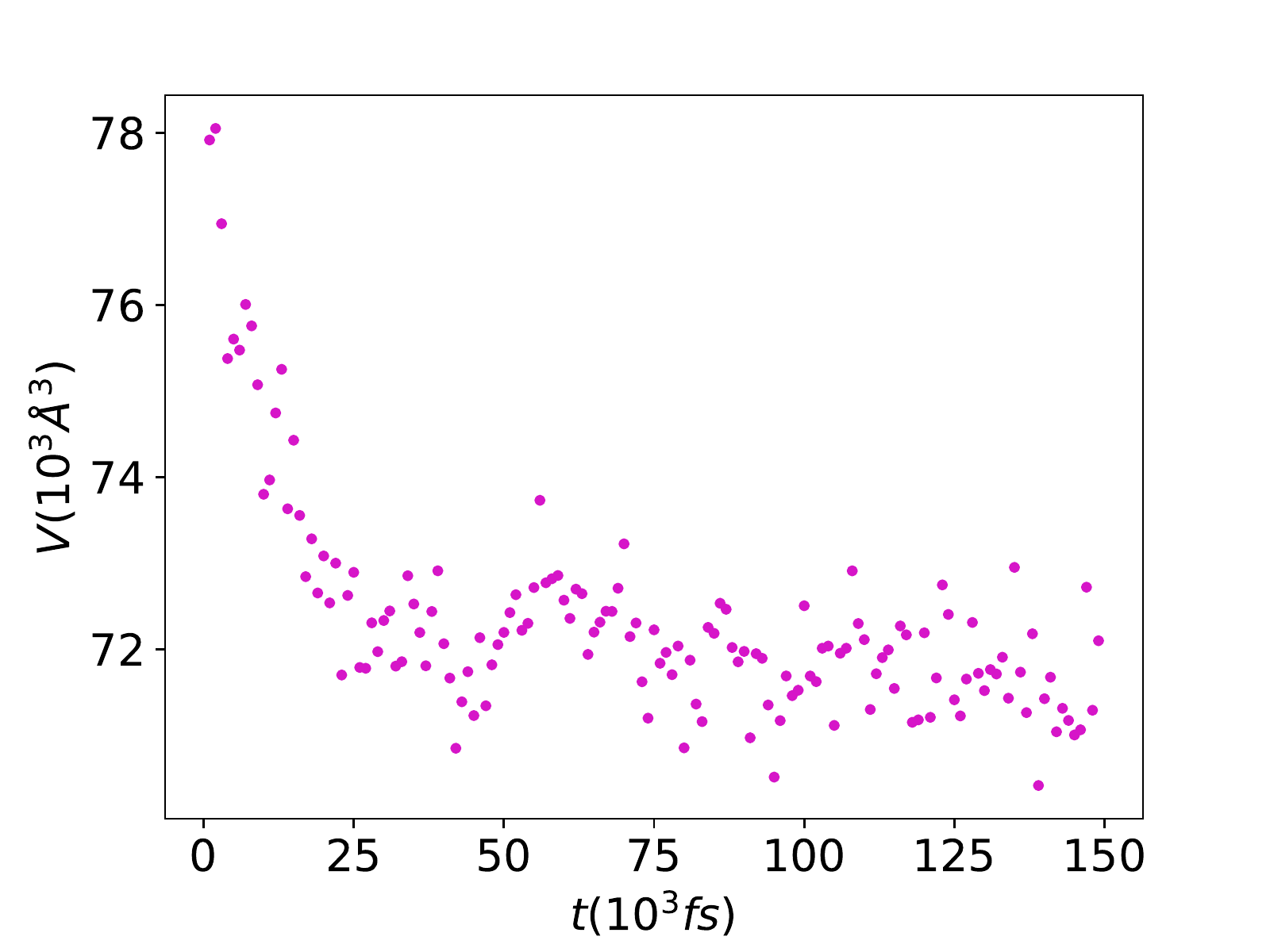}
	\caption{Volume vs simulation time for density of 4-ethyl-4-methylheptane at $25^{\circ}$C for first 150ps of the simulation. $V(t=0)=64000$\AA$^3$ and is not shown since it's a guess system volume. System equilibriates after approximately 25000 time steps, but we conservatively consider volume measurements only after 30000 time steps.}
	\label{fig:eqtime}
\end{figure}

To calculate the mean value in density and its uncertainty, first 30000 time steps are discarded to take the measurements only after the system is equilibriated (\autoref{fig:eqtime}). From the subsequent time steps, the expected value of density is calculated as a ratio of cell mass and mean cell volume.

\subsection{Data blocking}\label{2.2}
Since the motion of atoms and molecules during the molecular dynamics simulations is deterministic, consecutive measurements of physical quantities are correlated, which results in underestimating their uncertainty.

To accurately determine the uncertainty in the property of interest, we use data blocking \cite{13} \space \cite{28}. In data blocking, consecutive measurements are first assembled into blocks of equal size. Next, the mean of each block is taken as its representative value. The uncertainty in the property of interest is calculated as a standard deviation in the mean in the blocked set. To obtain an actual value of uncertainty, blocking procedure is performed iteratively until the uncertainty reaches its maximum. In this manuscript, the number of data entries is halved with each blocking round. 

\begin{figure}[H]
	\centering
	\includegraphics[width=0.55\linewidth]{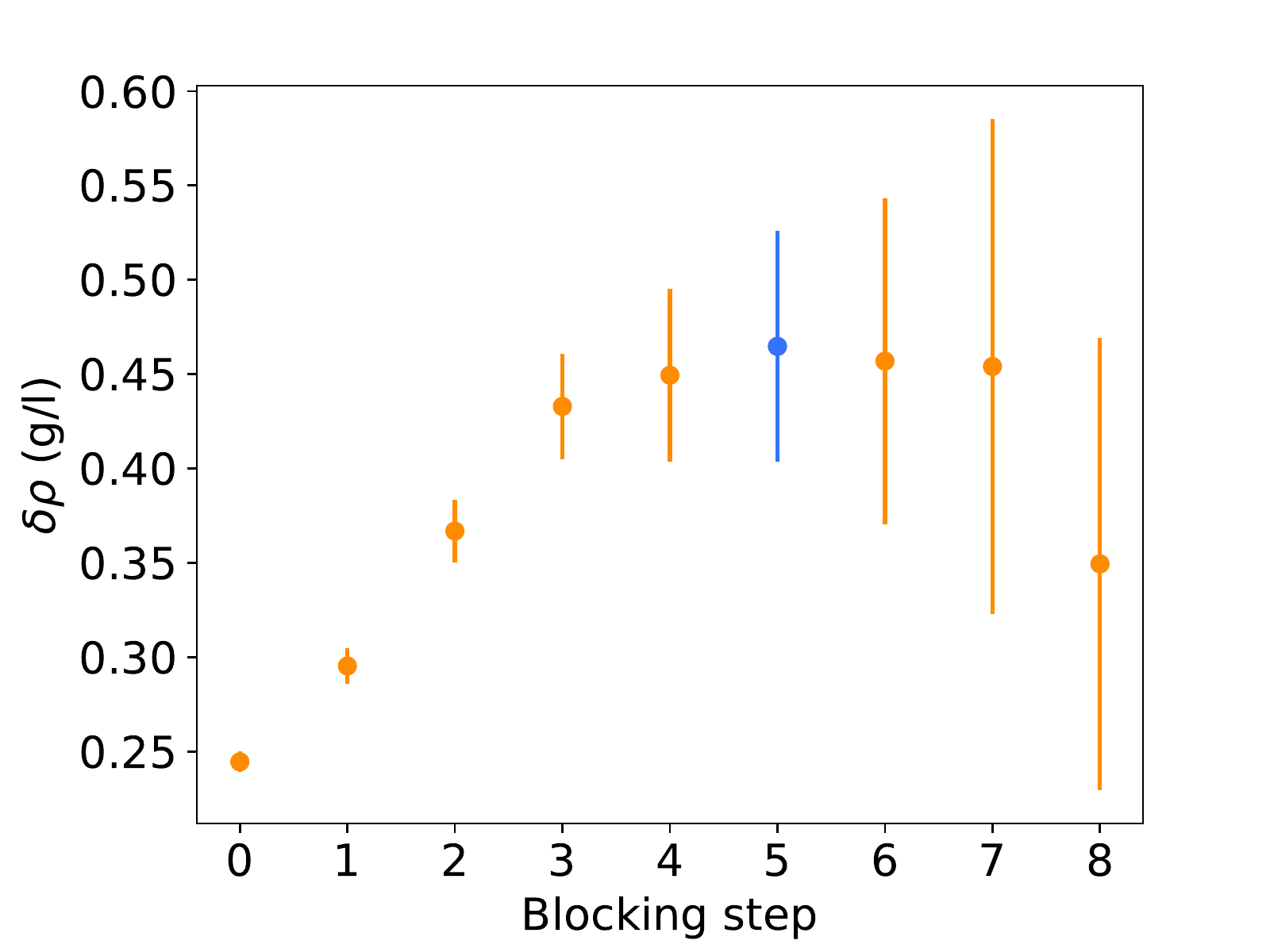}
	\caption{Applying data blocking to determine the uncertainty in density of 4,4-dimethyl-heptane at 100\degree C. Dots represent expected values of uncertainties while bars represent their uncertainties.}
	\label{fig:blocking}
\end{figure}

We illustrate the data blocking procedure by determining the uncertainty in density of 4,4-dimethyl-heptane at 100$^{\circ}$C (\autoref{fig:blocking}), which arises from expansion and contraction of the simulation cell. In this example, volume measurements become uncorrelated after five blocking rounds (highlighted by a blue dot with error bars in \autoref{fig:blocking}). 

\subsection{Details of molecular dynamics viscosity simulations}\label{2.3}
Molecular dynamics is frequently employed to simulate the Couette flow and determine viscosity of a liquid. To model viscosity, a system of alkane molecules in the liquid phase is trapped between two parallel infinite plates. Shear along the xy plane is then applied to one of the plates to make it move with a constant speed relative to the stationary plate. Consequently, alkane molecules move with the horizontal velocity component proportional to their vertical distance from the other plate.

Preparing input files to perform viscosity simulation comprises the same steps to perform density simulations, but with seven adjustments. Firstly, the density of the simulation cell is either the experimental or the average predicted density of alkane at the temperature of interest. Secondly, the xy component of the stress tensor ($P_{xy}$) is recorded every 100 time steps. Thirdly, since uncertainty in kinematic viscosity is inversely proportional to the shear rate, simulation time varies as a function of the shear rate at which the simulation is performed (\autoref{tab:SR}) to improve the confidence in viscosity predictions.  Fourthly, as the system is not kept at a constant pressure, no barostat is applied. Fifthly, simulations are performed in the NVT ensemble with the SLLOD\cite{SLLOD} equations of motion and the Lees-Edwards\cite{Lees-Edwards} boundary conditions. Sixthly, the velocity of each atom is rescaled during the simulation if the system temperature deviates from the initial temperature by more than a 100K. Finally, atoms do not exert a force on each other if the distance between them is smaller than 0.2 distance units. 

\begin{table}[H]
	\begin{center}
		\begin{tabular}{|c|c|} % <-- Alignments: 1st column left, 2nd middle and 3rd right, with vertical lines in between
			\hline
			\boldmath{$\rm log(\dot{\gamma})$}\space \boldmath{$(\rm s^{-1}$)}& \boldmath{$\rm T_{sim}$}\space \textbf{(ns)} \\ \hline
			10.30-12.00 & 1 \\ \hline
			10.10-10.30 & 2 \\ \hline
			9.35-10.10 & 4  \\ \hline
			$<$9.35 & 8 \\ \hline
		\end{tabular} 
		\caption{Viscosity simulation time as a function of shear rate.}
		\label{tab:SR}
	\end{center}
\end{table}

In experiments, viscosity measurements are performed at constant pressure. However, since density is also constant during viscosity experiments, use of the NVT ensemble is physically justified. Simulations can also be performed in the NPT ensemble, but we decide to perform them in the NVT ensemble since "barostats (which alter positions through volume changes) greatly affect the dynamics of the system" \cite{2} and it's more commonly applied.

The expected value of kinematic viscosity at shear rate $\dot{\gamma}_{i}$ is calculated as the ratio of the negative expected value of the xy component of the viscous stress tensor and a product of the simulation shear rate and liquid density

\begin{equation}
\eta(\dot{\gamma}_{i})=\frac{-\E[P_{xy}]}{\rho\dot{\gamma}_{i}},
\end{equation}

where $\E$, $\rho$ and $\dot{\gamma}_{i}$ denote the expectation operator, alkane's density and the shear rate. Uncertainty in kinematic viscosity is calculated as the ratio of the uncertainty in the xy component of the viscous stress tensor and a product of shear rate and liquid density

\begin{equation}
\delta\eta(\dot{\gamma}_{i})=\frac{\delta P_{xy}}{\rho\dot{\gamma}_{i}},
\end{equation}

where $\delta P_{xy}$ denotes the uncertainty in the xy component of the shear stress tensor and we have neglected the uncertainty in density, since typically $\frac{\delta{\rho}}{\rho}\ll\frac{\delta P_{xy}}{P_{xy}}$. Uncertainty in kinematic viscosity can in principle be reduced by increasing the simulation time, but its ultimate minimum is in practice limited by its dependence on the reciprocal shear rate. 

To calculate Newtonian viscosity, viscosity's shear rate profile is fitted to the Carreau model

\begin{equation}
\eta(\dot\gamma)=\eta_{\infty}+(\eta_{0}-\eta_{\infty})[1+(\lambda\dot\gamma)^2]^{\frac{n-1}{2}},
\end{equation}

where $\eta_{0}$ and $\eta_{\infty}$ are the values of upper and lower Newtonian plateaus, $n$ is a nonnegative parameter that determines the shape of the Carreau curve between two plateaus, and $\lambda$ determines the range of shear rates between the two plateaus. To calculate Newtonian viscosity, we minimize the weighted least squares (WLS) cost function

\begin{equation}
C(\eta_{0},\eta_{\infty},n,\lambda)=\sum_{i}\frac{[\eta_{i}-\eta({\gamma_{i}})]^2}{\delta\eta_{i}^2},
\end{equation}

where $\eta_{i}$ is the simulation result of kinematic viscosity at $\gamma_{i}$,$\eta({\gamma_{i}})$ is the Carreau model viscosity at shear rate $\gamma_{i}$, and $\delta\eta_{i}$ is the uncertainty in kinematic viscosity. The choice of cost function ensures that we assign higher weights to viscosity results at higher shear rates, with a larger signal to noise ratio to ensure a more reliable extrapolation of Newtonian viscosity. We minimize the WLS cost function with the Levenberg-Marquardt algorithm \cite{Lev1} \space \cite{Lev2} and the initial parameter guesses of

\begin{equation}
\{ \eta_{0},\eta_{\infty},n,\lambda\}=\bigg\{{\rm max\{\eta\},min\{\eta\},1,\frac{1}{min\{\dot\gamma\}}}\bigg\},
\end{equation}

where $\{n\}$ and $\{\gamma\}$ are the set of viscosity simulation results and shear rates at which we perform the simulations.

\subsection{Identifying good shear rates}\label{2.4}
The ratio of speeds due to shear and due to particle interactions is proportional to the shear rate, resulting in a low signal to noise ratio for viscosity simulations performed at low shear rates. For a fixed shear rate, this ratio is smaller for larger temperatures due to smaller relative contribution of the kinetic term to the shear stress tensor, heavier molecules due to inverse relation between speed at a fixed temperature and molecular mass, and at higher pressures due to an increased virial term contribution arising from closer proximity of molecules at a fixed volume. Therefore, direct identification of Newtonian viscosity with NEMD is challenging, while poor statistics at low shear rates becomes an obstacle in its accurate extrapolation. The range of good shear rates at which to perform viscosity simulations is a priori unknown, and while the authors of previous NEMD studies have performed their simulations at reasonable shear rates, they have selected them manually. Currently, an algorithm to automatically sample good shear rates for an arbitrary alkane at any temperature and pressure does not exist.

\begin{figure}[H]
	\centering
	\includegraphics[width=0.8\linewidth]{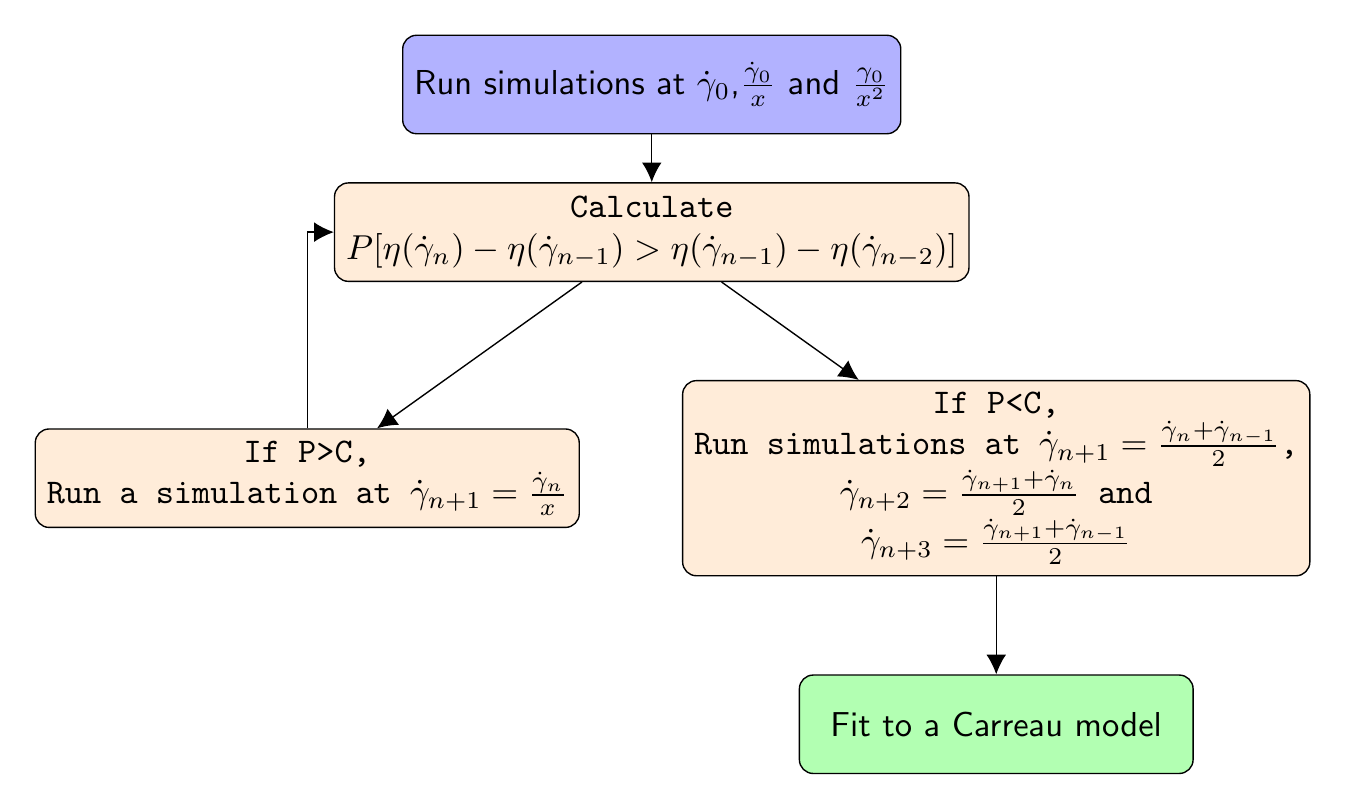}
	\caption{Schematic of the algorithm applied to determine appropriate shear rates.}
	\label{fig:flowchart}
\end{figure}

To automatically sample good shear rates for an arbitrary alkane, we first run a simulation at the largest shear rate $\dot{\gamma}_{0}$. Next, we successively decrease the shear rate by a constant $x>1$ and perform simulations at two smaller shear rates $\dot{\gamma}_{1}=\frac{\dot{\gamma_{0}}}{x}$ and $\dot{\gamma}_{2}=\frac{\dot{\gamma_{0}}}{x^2}$. Then, to assess the vicinity to the upper Newtonian plateau, we calculate the probability that the shear rate profile of kinematic viscosity between two smallest shear rates is concave up,  $P[\eta(\dot\gamma_{2})-\eta(\dot\gamma_{1})>\eta(\dot\gamma_{1})-\eta(\dot\gamma_{0})]$ and compare it to a constant $C \in [0,1]$ under the assumption that kinematic viscosity at each shear rate is normally distributed under its mean and the uncertainty. If $P[\eta(\dot\gamma_{2})-\eta(\dot\gamma_{1})>\eta(\dot\gamma_{1})-\eta(\dot\gamma_{0})]>C$, we again decrease the shear rate by a constant $x$ and run a simulation at $\dot{\gamma}_{3}=\frac{\dot{\gamma_{2}}}{x}$ before we determine the probability that viscosity's shear rate profile between $\dot{\gamma}_{1}$ and $\dot{\gamma}_{3}$ is concave up. The process of performing the simulations at successively smaller  shear rates that are a constant fraction of the previous shear rate is repeated until $P[\eta(\dot\gamma_{n})-\eta(\dot\gamma_{n-1})>\eta(\dot\gamma_{n-1})-\eta(\dot\gamma_{n-2})]<C$. To avoid performing simulations with a low signal to noise ratio, we do not perform the simulations at smaller shear. Instead, we perform three more simulations at shear rates uniformly spaced between two smallest shear rates $\dot{\gamma}_{n+1}=\frac{\dot{\gamma}_{n}+\dot{\gamma}_{n-1}}{2}$, $\dot{\gamma}_{n+2}=\frac{\dot{\gamma}_{n+1}+\dot{\gamma}_{n}}{2}$, and $\dot{\gamma}_{n+3}=\frac{\dot{\gamma}_{n+1}+\dot{\gamma}_{n-1}}{2}$. In this manuscript, we use $\dot{\gamma}_{0}=10^{12}\rm s^{-1}$, $x=3$ and $C=0.95$ to cover a large range of shear rates with a relatively small number of simulations and continue performing simulations at smaller shear rates only if we're 95\% confident that viscosity's shear rate profile in the region of interest is concave up. A flow chart that concisely summarizes the sampling algorithm is shown in \autoref{fig:flowchart}. 

\begin{figure}[H]
	\centering
	\includegraphics[width=0.55\linewidth]{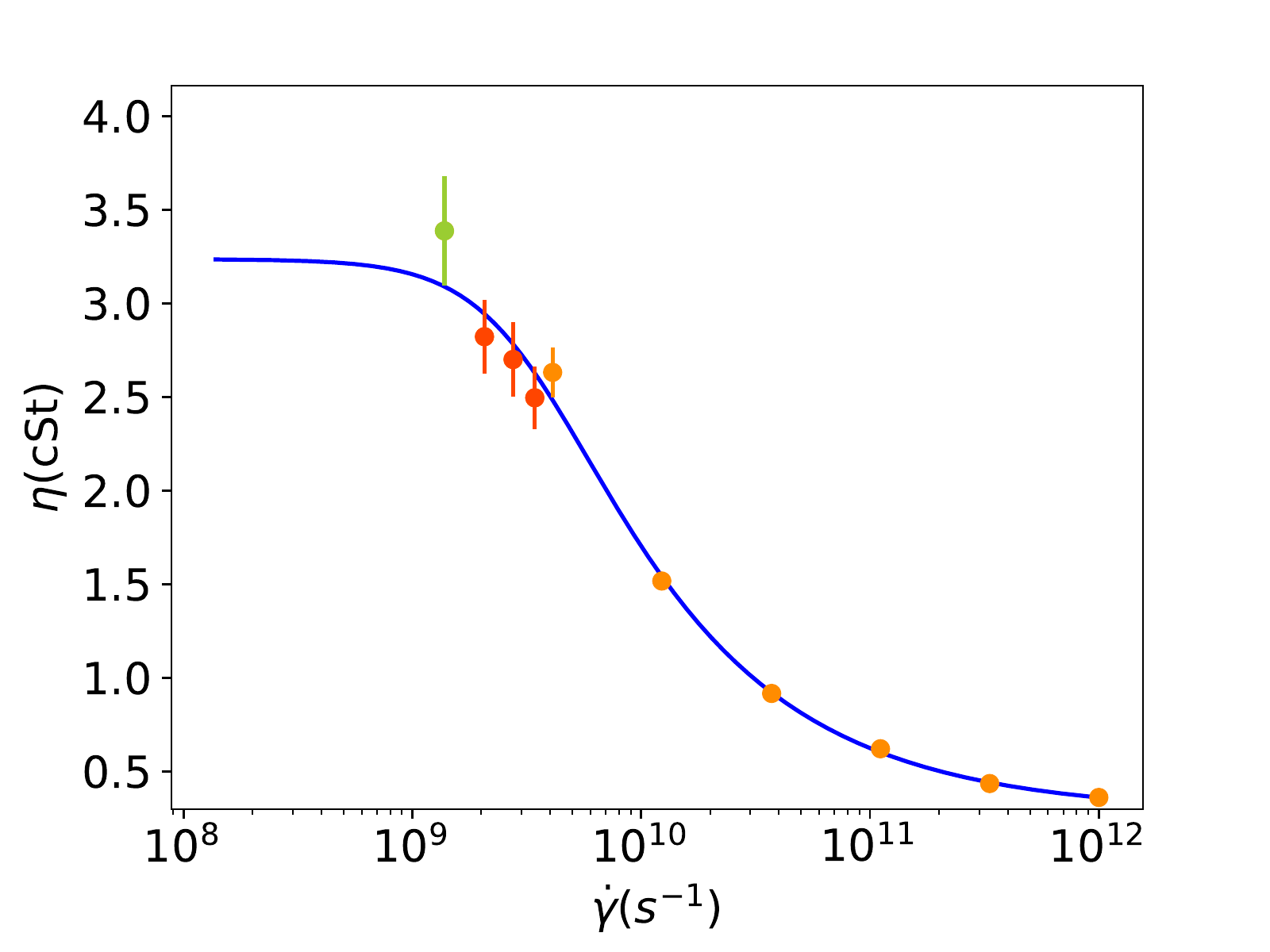}
	\caption{Kinematic viscosity plotted against the shear rate for octadecane at 50$^{\circ}$C. Shear rate is plotted on the logarithmic scale. Orange dots represent simulation results at high shear rates; light green dot represents the simulation result at the lowest shear rate, while red dots represent simulation results at intermediate shear rates.}
	\label{fig:algorithmexample}
\end{figure}

We illustrate the sampling algorithm in modelling kinematic viscosity of octadecane at 50\degree C (\autoref{fig:algorithmexample}). Shear rate is consecutively decreased by a third down to $\rm log(\dot{\gamma})=9.14$, when $\eta=3.39\pm0.29$cSt. Kinematic viscosity at two immediate smaller shear rates ($\rm log(\dot{\gamma})=9.61$ and $\rm log(\dot{\gamma})=10.09$) is simulated and found to be $2.63\pm0.13$cSt and $1.52\pm0.05$cSt. Since $P[\eta(10^{9.14})-\eta(10^{9.61})>\eta(10^{9.61})-\eta(10^{10.09})]=0.1847$, three more simulations are performed at $\rm log(\dot{\gamma})=9.44$, $\rm log(\dot{\gamma})=9.32$ and $\rm log(\dot{\gamma})=9.54$, and data is fitted to the Carreau model with a WLS regression (\autoref{2.3}).

While neither simulation has been performed at a low enough shear rates to directly identify the upper Newtonian plateau, we have extrapolated Newtonian viscosity of 3.24cSt, an excellent agreement with the experimental value of 3.23cSt reported in Caudwell et al. \cite{doi:10.1021/je800417q}.

\section{Results and Discussion}\label{3}
With molecular dynamics simulation technique in place, we are well-positioned to determine liquid density and kinematic viscosity of alkanes. First, we model density and compare to experimental values before repeating the process for kinematic viscosity.

In \autoref{3.1}, we model density of liquid alkanes with 5 to 10 carbon atoms and in \autoref{3.2} we model kinematic viscosity of hexane, heptane, octane, nonane, decane, undecane, dodecane, tridecane and tetradecane at 20\degree C, tridecane at 60$^{\circ}$C as a function of pressure, and octane, dodecane and octadecane as a function of temperature. To directly evaluate the performance of the sampling algorithm, we perform viscosity simulations at experimental densities (\autoref{tab:Linear}, \autoref{tab:TridecaneP}, \autoref{tab:OctaneT}, \autoref{tab:DodecaneT}, \autoref{tab:OctadecaneT}).

\subsection{Density}\label{3.1}
We perform molecular dynamics simulations in the NPT ensemble for 74 alkanes at 25\degree C and 34 alkanes at 100\degree C, and compare to experimental data from the TRC Thermodynamic Tables\cite{TRC}.

To check the reliability of NPT simulation results, we investigate the average pressures obtained in the NPT simulations. During a typical simulation, pressure varies between -1500atm and 1500atm. However, after averaging within a simulation, pressure varies between -30atm and 30atm, with statistical uncertainty obtained through data blocking (\autoref{2.2}) from between 20atm and 30atm, and the atmospheric pressure within the 95\% confidence interval. 

Next, we analyse the statistical uncertainty in density simulations. First, we compare uncertainties at two different temperatures. The average uncertainty at 25\degree C is 0.53 $\frac{g}{l}$, while at 100\degree C it is 0.61 $\frac{g}{l}$. There is no indication that increasing the temperature by 75\degree C increases the statistical uncertainty in density simulation results for small alkanes. Then, we investigate the uncertainty as a function of molecular weight (\autoref{tab:uncerho}) and conclude that increasing molecular weight does not increase the uncertainty in average densities for light linear, single-branched, and double-branched alkanes.

\begin{table}[H]
	\begin{center}
		\begin{tabular}{|c|c|c|c|} % <-- Alignments: 1st column left, 2nd middle and 3rd right, with vertical lines in between
			\hline
			\boldmath{$\rm N_{C}$}& \boldmath{$\rm N_{mol}$} & \boldmath{$\rm \E[\delta\rho]$}\space \boldmath{$(\rm \frac{g}{l}$)}\\ \hline
			6 & 10 & 0.69 \\ \hline
			7 & 16 & 0.51  \\ \hline
			8 & 26  & 0.57 \\ \hline
			9 & 24 & 0.58 \\ \hline
			10 & 32 & 0.50 \\ \hline
		\end{tabular} 
		\caption{Summary of uncertainty as a function of molecular weight. $\rm N_{C}$, $\rm N_{mol}$ and $\rm \E[\delta\rho]$ denote the number of carbon atoms, the total number of simulation results for molecules with the $ N_{C}$ number of carbon atoms and the mean value of uncertainty in density of alkanes with a fixed number of carbon atoms.}
		\label{tab:uncerho}
	\end{center}
\end{table}

\begin{figure}[H]
	\centering
	\begin{minipage}[a]{0.49\linewidth}
		\includegraphics[width=\linewidth]{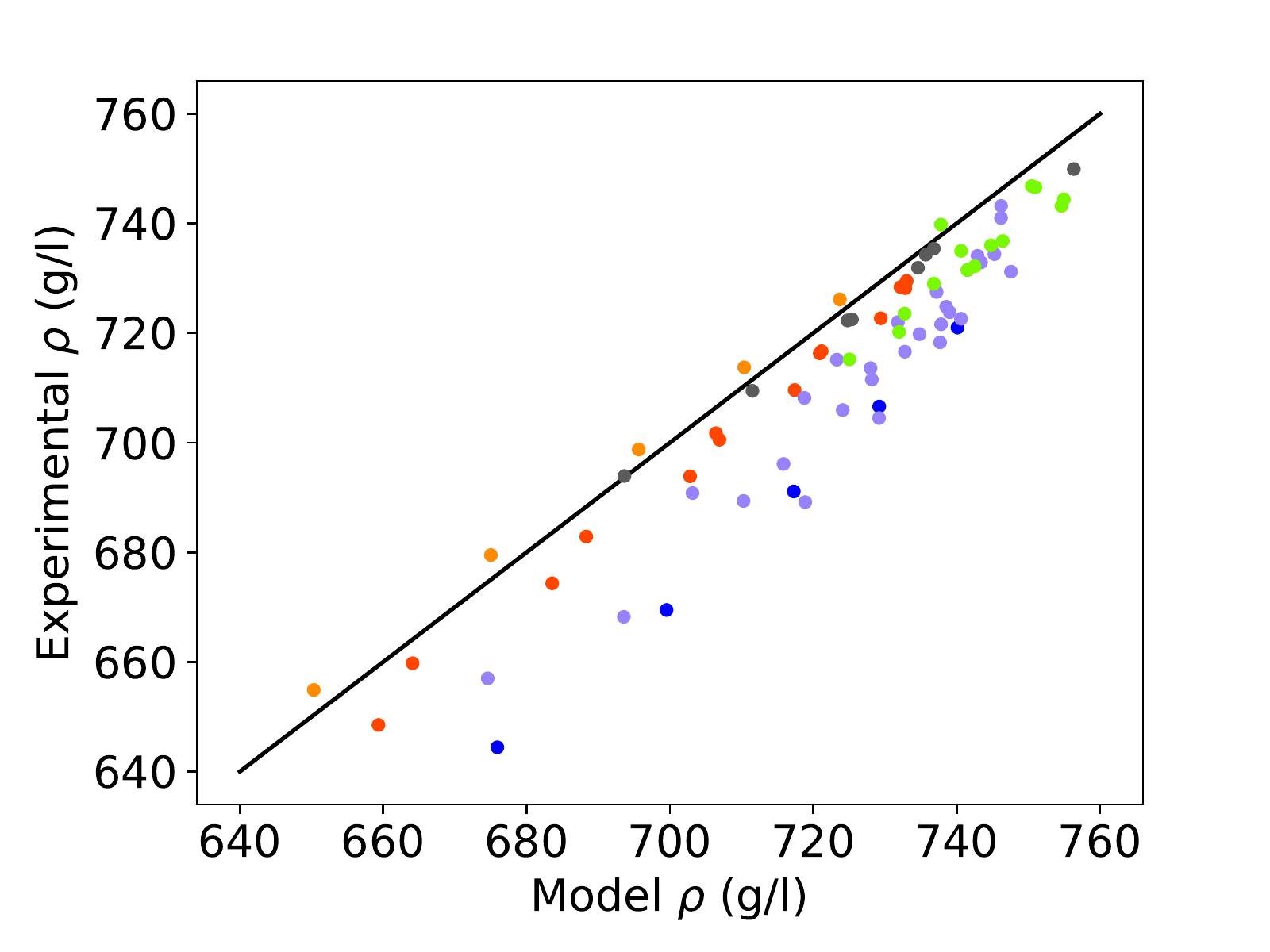}
	\end{minipage}
	\hfill
	\begin{minipage}[a]{0.49\linewidth}
		\includegraphics[width=\linewidth]{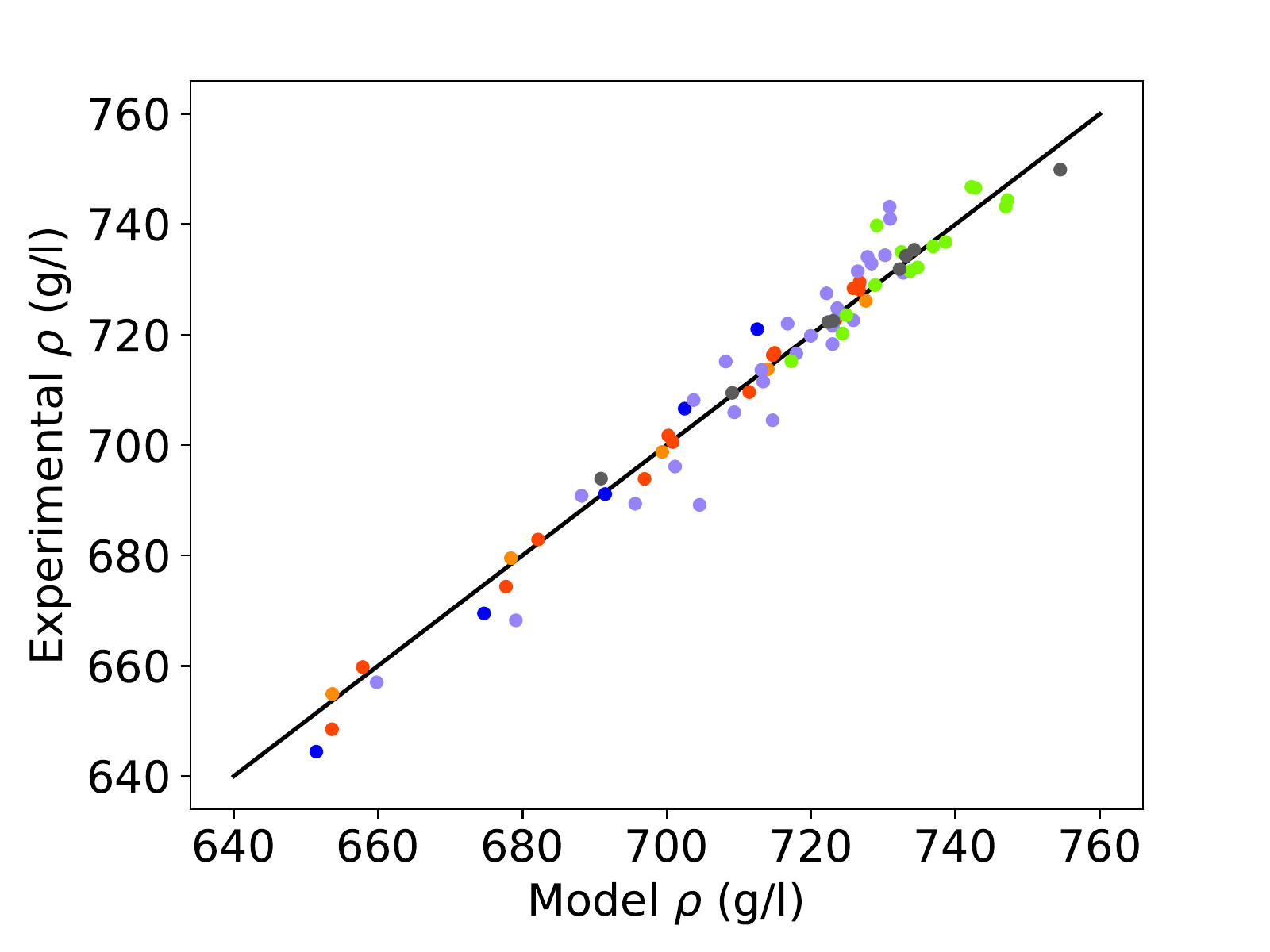}
	\end{minipage}
	\caption{Parity plot of density results vs experimental values before (left) and after (right) correction factors are applied. Orange dots denote the linear alkane series, red dots denote the methyl group, blue dots denote the 2,2-dimethyl series, violet dots denote the group comprising all the other alkanes, light green dots denote the ethyl-methyl group, while the group of all the other molecules is denoted with grey dots.}
	\label{fig:before_and_after}
\end{figure}

Initially, we obtain an average absolute deviation of 11$\frac{g}{l}$ at 25\degree C, and of 16$\frac{g}{l}$ at 100\degree C. To further understand the performance of density simulations, we drill into the discrepancy between experimental values and simulation results. We split the alkanes for which we performed simulations into six groups, with each group either a homologous series or a set of homologous series (\autoref{tab:errrho}). The discrepancy between simulations and experiment within each group is approximately constant (\autoref{fig:before_and_after}, left) , likely due to the systematic bias in the SciPCFF force field, as the density of alkanes with fewer branches is modelled more accurately, indicating that its parameters for alkanes have been developed mostly from linear alkane data. For each group apart from the linear alkanes, the average group discrepancy is larger at 100\degree C than at 25\degree C, possibly due to development of the SciPCFF force field parameters mostly from room temperature data.

\begin{table}[H]
	\begin{center}
		\begin{tabular}{|c|c|c|c|c|} % <-- Alignments: 1st column left, 2nd middle and 3rd right, with vertical lines in between
			\hline
			\textbf{Group}& \boldmath{$\rm N_{mol}$} \textbf{25 \degree C} & \textbf{$\left|\Delta_{25^{\circ}\rm C}\right|$} \boldmath{$(\rm \frac{g}{l}$)}& \boldmath{$\rm N_{mol}$} \textbf{100 \degree C}&\textbf{$\left|\Delta_{100^{\circ}\rm C}\right|$} \boldmath{$(\rm \frac{g}{l}$)}\\ \hline
			linear & 5  & 3.6  & 5 & 2.5 \\ \hline
			methyl series & 14 & 6.1 & 7 & 9.4 \\ \hline
			2,2-dimethyl & 5 & 26 & 3 & 34 \\ \hline
			other dimethyl & 28 & 15 & 12 & 22 \\ \hline
			methyl-ethyl series & 14 & 7.9 & 4 & 20 \\ \hline
			other & 8 & 2.4 & 3 & 5.2 \\ \hline
		\end{tabular} 
		\caption{Summary of group discrepancies. At both temperatures, number of molecules and the average absolute deviation and standard deviation in discrepancy are presented. Since the signs of discrepancies are consistent for each group, for all the groups but the linear group, $\Delta=\left|\Delta\right|$, with $\Delta=-\left|\Delta\right|$ for linear alkanes.}
		\label{tab:errrho}
	\end{center}
\end{table}

Since viscosity simulations are in general performed at a constant density, results of NPT simulations with large discrepancies are insufficiently accurate to be used as state points for NVT simulations. To obtain more accurate results, we subtract the value of average discrepancy between a group to which an alkane belongs from the simulation result (\autoref{tab:errrho}). A small average pressure variation in simulation results justifies applying the same correction factor at all pressures, since the isothermal compressibility factor is approximately constant for the range of average pressures obtained from simulations \cite{Compr}. However, since applying correction factors to simulation results is a poor indication of the actual merit of applying them, which is why we perform a leave-one-out cross-validation \cite{StatisticalLearning}, in which correction factors are calculated from all but one data entry and applied to the remaining data entry, repeating for each entry in a dataset.

After applying a leave-one-out cross-validation, we obtain an average absolute deviation of 3.4 $\frac{\rm g}{\rm l}$ at 25\degree C (\autoref{fig:before_and_after}) and of 7.2 $\frac{\rm g}{\rm l}$ at 100\degree C, a significant improvement over the results obtained from molecular dynamics simulations. The summary of all the results is presented in \autoref{tab:errafter}, while the parity plot of corrected densities is presented in the right part of \autoref{fig:before_and_after}. At 25\degree C, the model performs the best for linear alkanes and the worst for the 2,2-dimethyl homologous series and the group of other dimethyl alkanes. At 100\degree C, the model still performs the best for linear alkanes, but now it performs the worst for the methyl-ethyl group, for which the average absolute deviation is 17 $\frac{\rm g}{\rm l}$. Such a large discrepancy arises from a large spread in discrepancies in original simulation results across the ethyl-methyl group. A full list of results can be found in \autoref{tab:density 25} and \autoref{tab:density 100}.

\begin{table}[H]
	\begin{center}
		\begin{tabular}{|c|c|c|c|c|} % <-- Alignments: 1st column left, 2nd middle and 3rd right, with vertical lines in between
			\hline
			\textbf{Group}& \boldmath{$\rm N_{mol}$} \textbf{25 \degree C} & \textbf{$\left|\Delta_{25^{\circ}\rm C}\right|$} \boldmath{$(\rm \frac{g}{l}$)}& \boldmath{$\rm N_{mol}$} \textbf{100 \degree C}&\textbf{$\left|\Delta_{100^{\circ}\rm C}\right|$} \boldmath{$(\rm \frac{g}{l}$)}\\ \hline
			linear & 5  & 0.93 & 5 & 0.84\\ \hline
			methyl series & 14 & 2.1 & 7 & 4.3 \\ \hline
			2,2-dimethyl & 5 & 5.0 & 3 & 5.4 \\ \hline
			other dimethyl & 28 & 4.9 & 12 & 9.6\\ \hline
			methyl-ethyl series & 14 & 3.1 & 4 & 17 \\ \hline
			other & 8 & 1.4 & 3 & 2.8 \\ \hline
		\end{tabular} 
		\caption{Summary of discrepancies after applying the correction factors and running a leave-one-out cross validation. At both 25 \degree C and 100 \degree C, number of molecules and absolute average deviation are presented.}
		\label{tab:errafter}
	\end{center}
\end{table}

Once liquid density at two temperatures is calculated, it is straightforward to determine it at any other temperature in the liquid phase due to its linear dependence on temperature.

\subsection{Viscosity}\label{3.2}
We now study kinematic viscosity of linear alkanes, first as a function of molecular weight, then as a function of pressure, and finally as a function of temperature. Linear alkanes serve as a case study for evaluating the reliability and accuracy of the sampling algorithm for two reasons. Firstly, they are the homologous alkane series with readily available experimental data. Secondly, systematic error in the SciPCFF force field for linear alkanes is likely small compared to the systematic error for the other homologous series. Consequently, the discrepancy between simulations and the experiments arises primarily from the remaining noise in viscosity simulations.

\begin{figure}[H]
	\centering
	\includegraphics[width=0.55\linewidth]{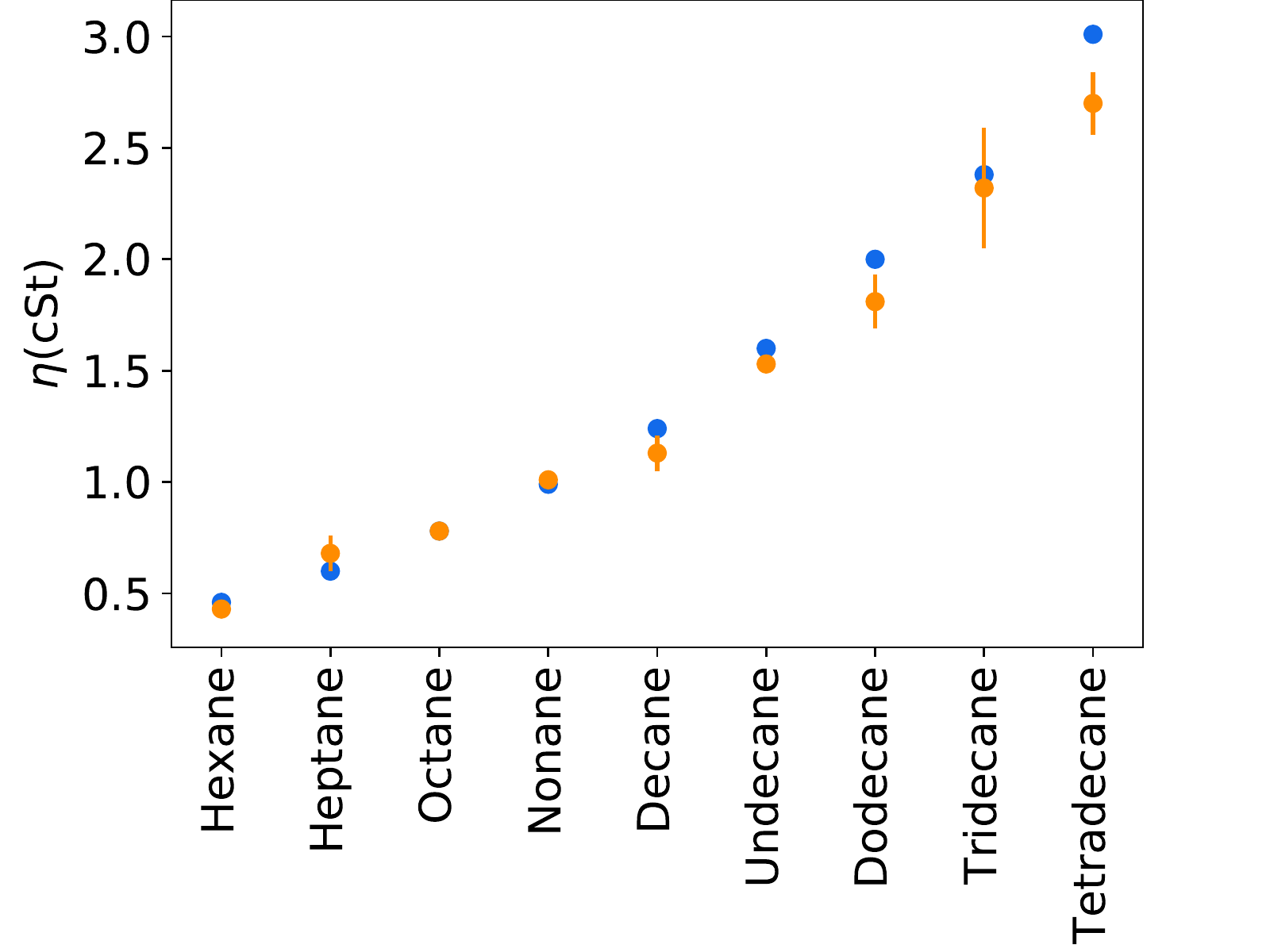}
	\caption{Viscosity of linear alkanes at 20\degree C. Blue dots present experimental data, while orange dots represent molecular dynamics predictions with accompanying statistical uncertainty.}
	\label{fig:ViscLinear}
\end{figure}

We first study viscosity as a function of molecular weight and model kinematic viscosity of hexane, heptane, octane, nonane, decane, undecane, dodecane, tridecane and tetradecane at 20\degree C at atmospheric pressure and compare results to experimental values from the TRC Thermodynamic Tables \cite{TRC} (\autoref{tab:Linear}). Simulations accurately reproduce the experimental data, with an average percent error of 5\% (\autoref{fig:ViscLinear}) and the absolute percent error of 6.4\%. Simulations are the least accurate for heptane and tetradecane, with the percent errors of 13\% and -10\%, while experimental values for all the alkanes apart from tetradecane are within the 95\% confidence interval. Simulations systematically underestimate kinematic viscosity of decane and heavier alkanes, which we attribute to the small systematic error in the SciPCFF force field that also underestimated the density of linear alkanes.

To further evaluate the performance of the sampling algorithm, we compare the accuracy of our prediction for decane to the prediction made in Cui et al.\cite{1} at 25\degree C. Our prediction of 1.13$\pm$0.08cSt is in excellent agreement with the experimental value of 1.24cSt and compares favourably with their prediction of 0.84$\pm$0.11cSt against the experimental value of 1.17cSt.

\begin{figure}[H]
	\centering
	\includegraphics[width=0.55\linewidth]{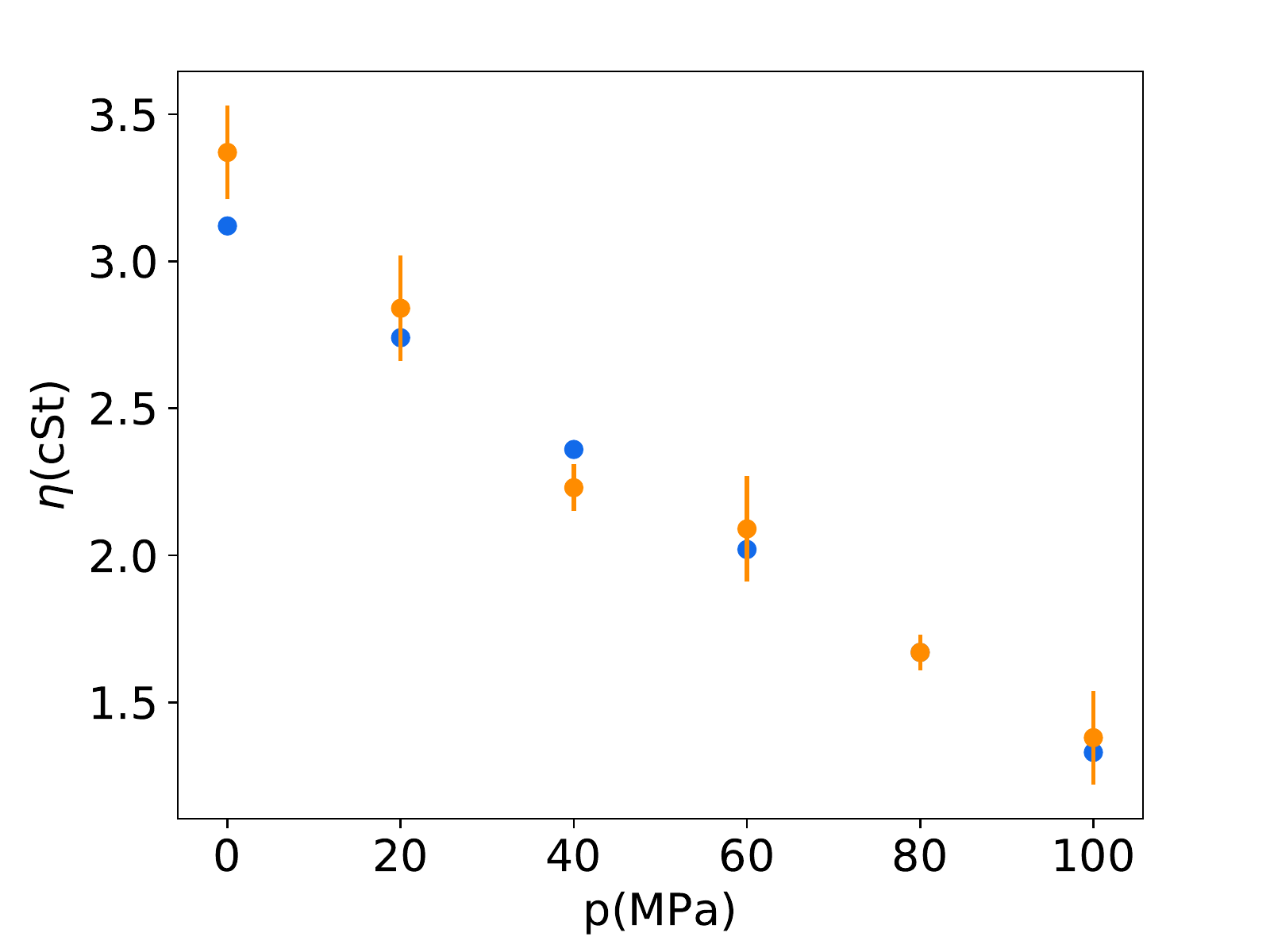}
	\caption{Viscosity of tridecane at 60\degree C as a function of pressure. Blue dots present experimental data, while orange dots represent molecular dynamics predictions with accompanying statistical uncertainty.}
	\label{fig:Tridecane}
\end{figure}

Secondly, we explore the variation of viscosity with pressure, with tridecane at 60\degree C as a case study and the experimental data coming from Daug et al.\cite{TridecanePressure} (\autoref{tab:TridecaneP}). Simulations results are in excellent agreement with experiments (\autoref{fig:Tridecane}), with an average percent error of 2\%, an absolute percent error of 4\%, and the least accurate prediction at 100MPa, with a percent error of 8\%. All the experimental values are within a 95\% confidence interval of our predictions.

Next, we calculate the pressure-viscosity coefficient, which is a measure commonly used in industry to assess the pressure gradient of alkanes' viscosity at a fixed temperature $T$. The pressure-viscosity coefficient appears in the exponent of the following equation:

\begin{equation}
\eta(p,T)=\eta_{\rm atm}(T)\rm e^{\alpha p},
\end{equation} 

where $\eta_{\rm atm}(T)$ is a value of kinematic viscosity at atmospheric pressure and the temperature of interest, and p is the pressure. An experimental value of the pressure viscosity coefficient is 0.00886 MPa$^{-1}$, while the simulations predict 0.00869 MPa$^{-1}$. A percent error of only -2\% and the absolute percent error of 4\% further confirm that we can accurately capture the variation of alkane's viscosity with pressure.

Thirdly, we study the variation of viscosity with temperature, focusing on viscosity of octane, dodecane, and octadecane. Simulations are performed at temperatures at least 20\degree C above alkanes' melting points to avoid the crystallization of the cell.  

\begin{figure}[H]
	\centering
	\includegraphics[width=0.55\linewidth]{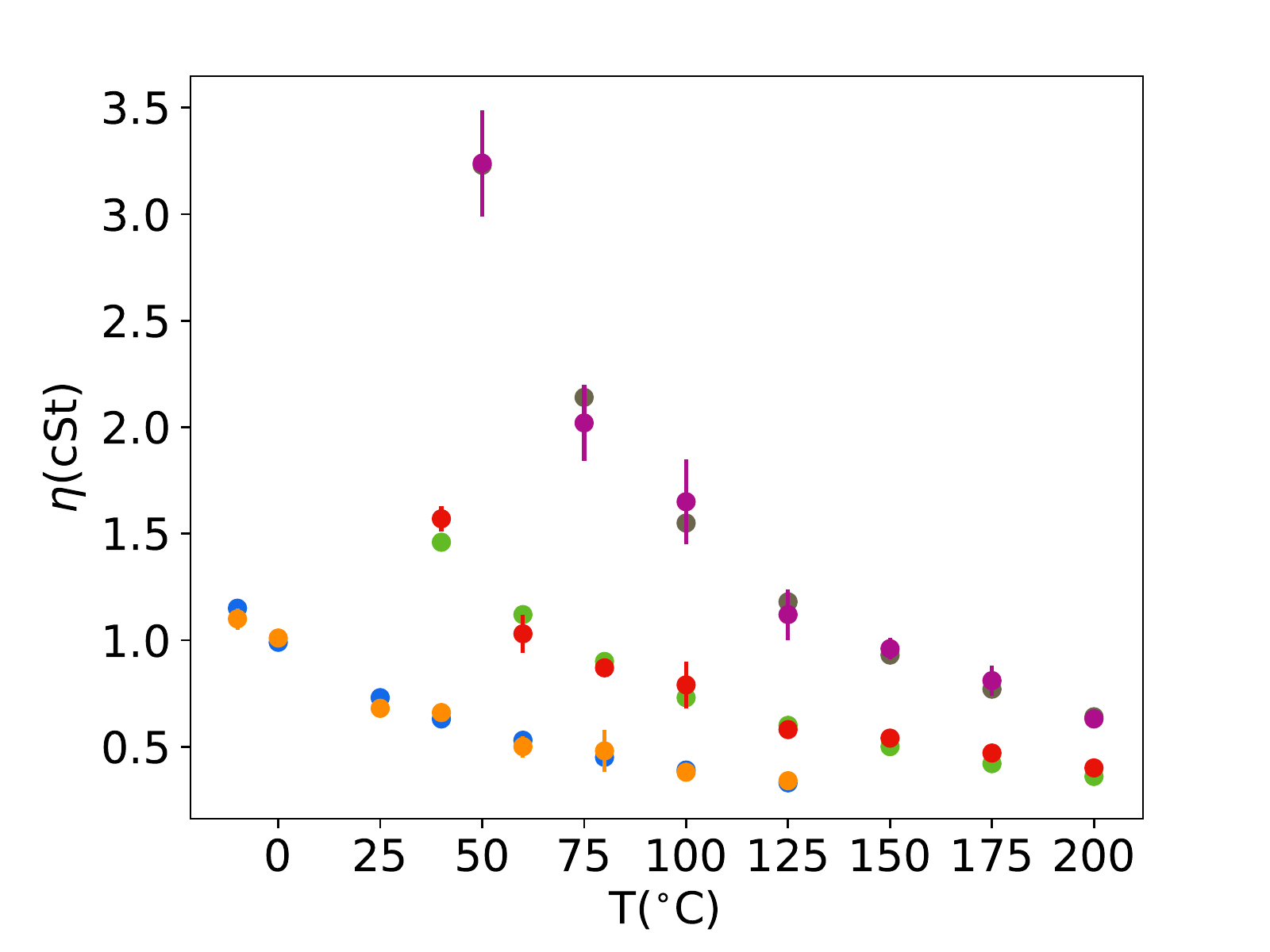}
	\caption{Viscosity of octane, dodecane and octadecane as a function of temperature. Blue, green and grey dots represent experimental values of their viscosity, while orange, red and purple dots with accompanying statistical uncertainty represent values predicted by the NEMD simulations.}
	\label{fig:OctDodOct}
\end{figure}

We first model viscosity of octane (\autoref{tab:OctaneT}) and dodecane (\autoref{tab:DodecaneT}), whose experimental viscosity's temperature profile was obtained from Caudwell et al. \cite{doi:10.1021/je800417q} \space \cite{Caudwell2004}. Simulation results are in excellent agreement with experiments, with the average percent error of -0.4\% for octane and of 4\% for dodecane, and the absolute percent error of 4\% and of 8\% for dodecane (\autoref{fig:OctDodOct}). All the experimental values lie within the 95\% confidence interval of mean simulation predictions apart from the octane results at 25 \degree C and 100 \degree C and the dodecane results at 200\degree C, primarily due to an excellent fit of viscosity's shear rate profile to the Carreau model.

Next, we study viscosity of octadecane, whose experimental values were obtained from Caudwell et al.\cite{Caudwell2004}. Simulations are in excellent agreement with experimental values (\autoref{fig:OctDodOct}), with an average deviation of 0.4\%, and the absolute average percent error of 4\%. Viscosity at 100\degree C was simulated with the smallest accuracy, with a 6\% percent deviation, while all the results apart from the one at 200\degree C are within a 95\% confidence interval. The longest total simulation time to model viscosity at a fixed temperature is 36ns, which is only 5.14 times longer than the time spent to model viscosity of hexane at 20\degree C. Such a small increase in total simulation time gives us further confidence that we can apply the sampling algorithm to heavy alkanes without requiring excessive computational resources like in equilibrium molecular dynamics.

\begin{figure}[H]
	\centering
	\includegraphics[width=0.55\linewidth]{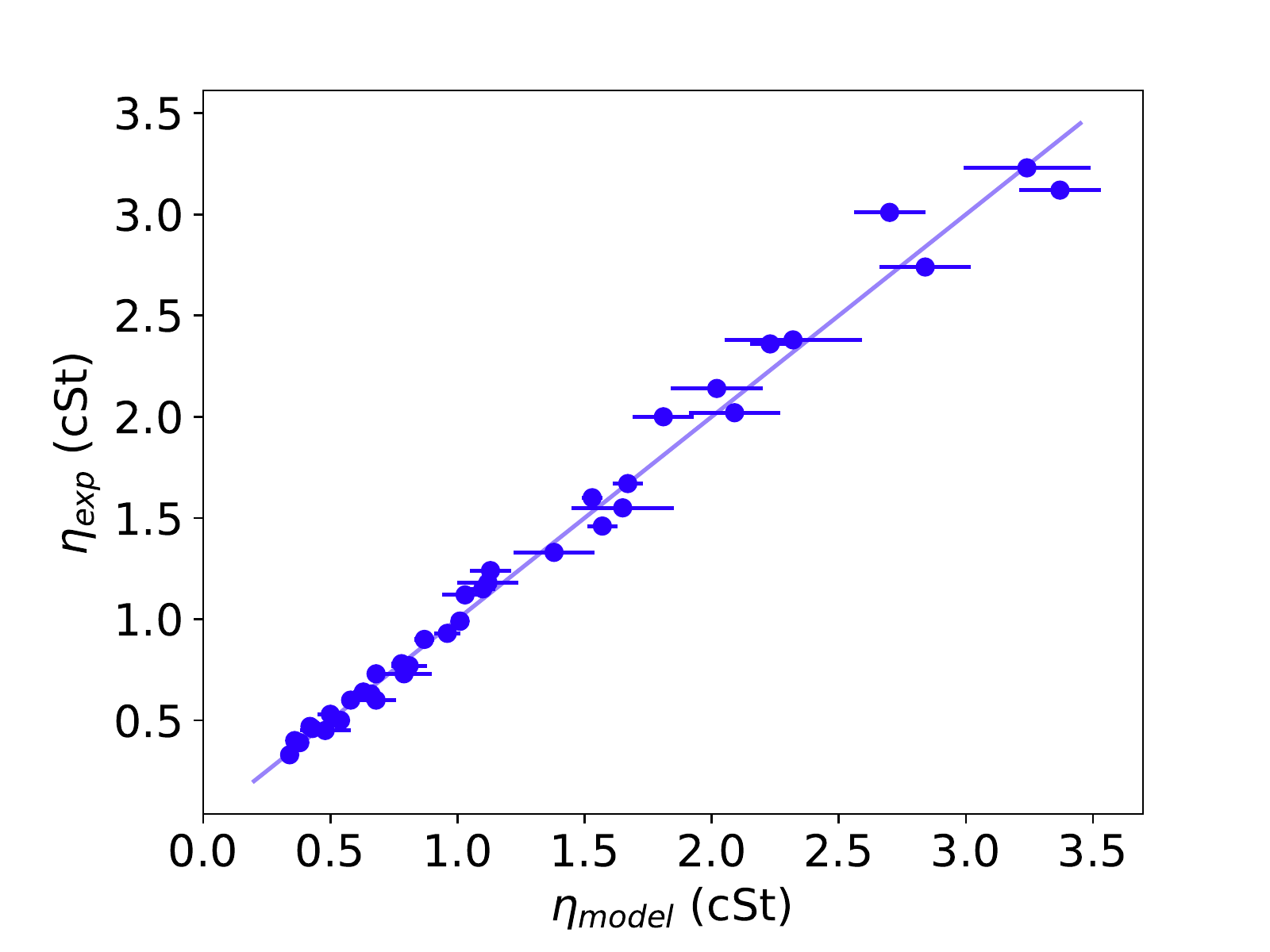}
	\caption{Parity plot showing experimental viscosity values against the NEMD simulation results.}
	\label{fig:ParityPlot}
\end{figure}

Having studied viscosity of linear alkanes as a function of pressure, temperature and molecular mass, we analyse the overall accuracy of viscosity simulations. A parity plot showing experimental values against simulation results for all the alkanes studied is shown in \autoref{fig:ParityPlot}. Simulations are in excellent agreement with experiments, with an average error of -1\% and the average absolute percent error of 5\%.

\begin{figure}[H]
	\centering
	\includegraphics[width=0.55\linewidth]{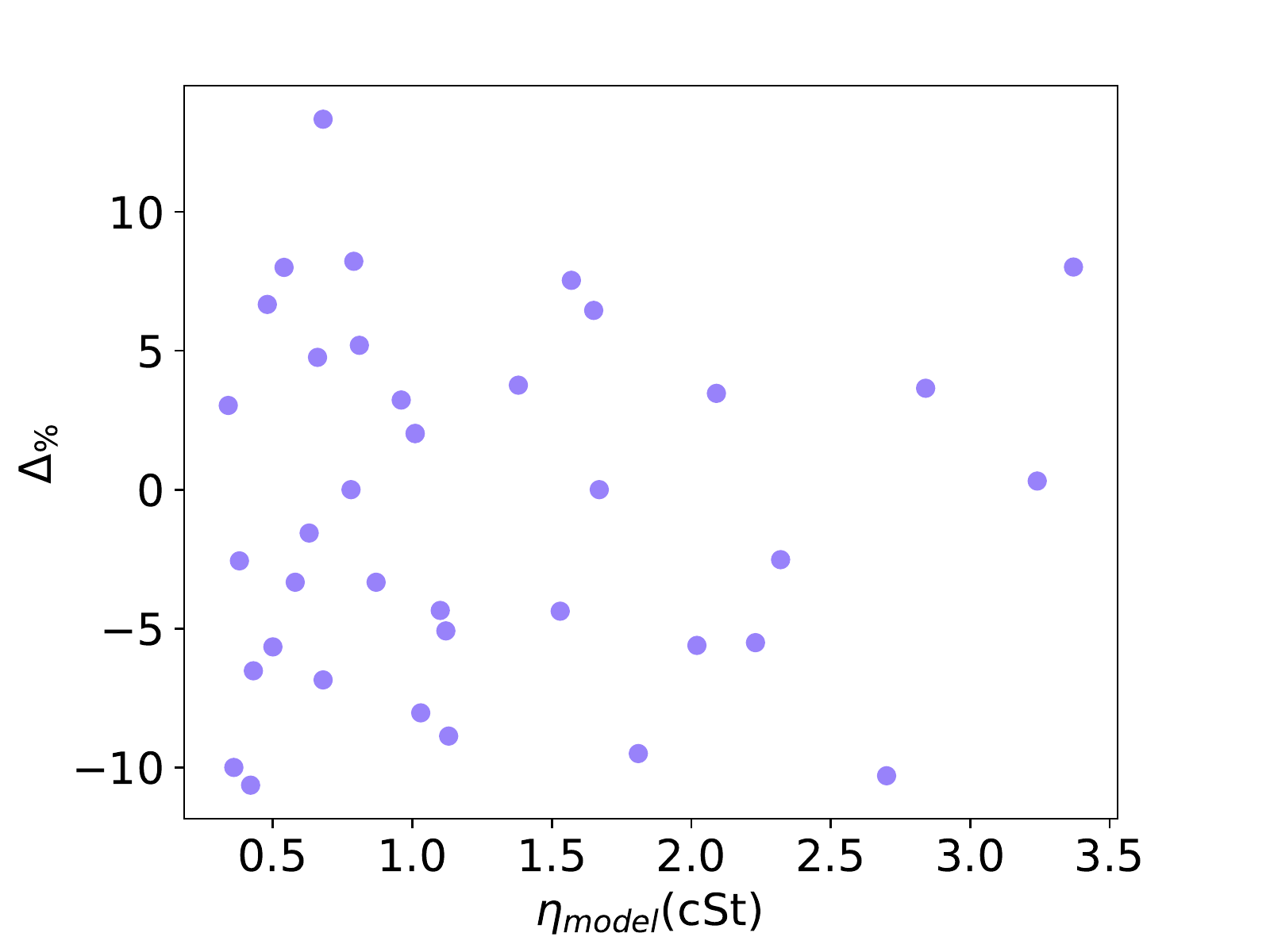}
	\caption{Percent error for all the data as a function of viscosity simulation results.}
	\label{fig:PercError}
\end{figure}

Then, we study the percent error in our models as a function of predicted viscosity to assess whether the simulations perform equally well at all viscosities (\autoref{fig:PercError}). We note that the average error fluctuates between about -10\% and 10\% for all the modelled viscosities, showing that the sampling algorithm  could be successfully applied to heavy alkanes.

\begin{figure}[H]
	\centering
	\includegraphics[width=0.55\linewidth]{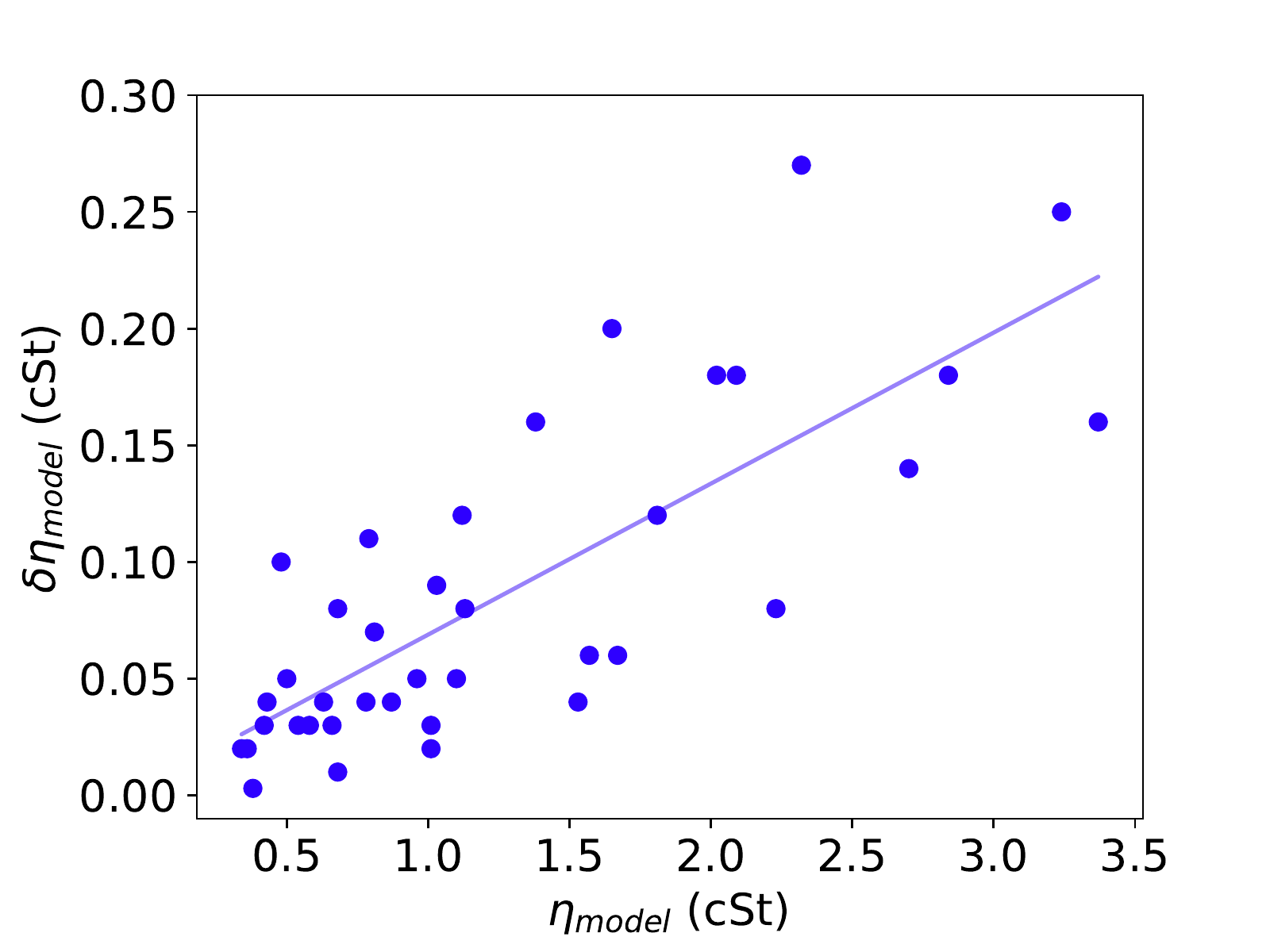}
	\caption{Uncertainty in Newtonian viscosity predictions as a function of viscosity predictions and its best linear fit $\delta\eta_{\rm model}=0.065\eta_{\rm model}+0.004$}
	\label{fig:AbsError} %ACTUALLY UNCERTAINTY AS F(PREDICTED VISCOSITY)
\end{figure}

Finally, we study the statistical uncertainty in the mean predictions. We observe that the uncertainty in NEMD viscosity predictions increases approximately linearly as a function of predicted viscosity (\autoref{fig:AbsError}), with an \rsq 0.61 of the linear fit. The approximate linear dependence of uncertainty on viscosity arises from uncertainty in the best fit parameters' dependence on the matrix of uncertainties in kinematic viscosity at different shear rates, whose entries are inversely proportional to the shear rate.

\section{Conclusion}\label{4}
In this manuscript, we have enhanced the existing molecular dynamics protocol to study liquid density and kinematic viscosity of alkanes. Firstly, we have studied density of alkanes with 5 to 10 carbon atoms by running simulations in the NPT ensemble and applied correction factors to simulation results to rectify the systematic error arising from the SciPCFF force field, obtaining an absolute deviation of 3.4 $\frac{g}{l}$ at 25\degree C and an absolute deviation of 7.2 $\frac{g}{l}$ at 100\degree C.

Secondly, we have also developed a sampling algorithm to identify the shear rates at which to perform viscosity simulations. We have utilised the sampling algorithm to study the kinematic viscosity of hexane, heptane, octane, nonane, decane, undecane, dodecane, tridecane, and tetradecane at 20\degree C; viscosity of tridecane at 60\degree C as a function of pressure, and viscosity of octane, dodecane, and octadecane as a function of temperature at experimental densities. Simulations are in excellent agreement with experiments, with an average percent error of -1\% and the average absolute percent error of 5\%. The average percent error stays approximately constant and fluctuates about 10\% in magnitude as a function of viscosity, while the uncertainty in viscosity predictions increases approximately linearly with increased viscosity.

Formalism presented in this manuscript sets a solid foundation to determine density and viscosity of larger and more complex alkanes. Collecting more experimental data and performing additional molecular dynamics simulations for density would enable us to further exploit systematic errors arising from the SciPCFF force field, while machine learning \cite{me} can be used to predict simulation results for various molecules without explicitly performing the simulations. 

The sampling algorithm that automatically determines shear rates can straightforwardly be applied in high throughput screening, while its generality means that it can be used as a basis to study viscosity of other liquids with a known functional dependence on shear rates. Mathematical properties of the sampling algorithm and the effects of using a multi-step/constraint algorithm in simulations can also be studied, so that the performance of NEMD NVT simulations approaches its optimum.

\section{Acknowledgments}
Pavao Santak would like to acknowledge the funding and the technical support from BP through the BP International Center for Advanced Materials (BP-ICAM), which made this research possible. Gareth Conduit would like to acknowledge financial support from the Royal Society. Both authors thank Leslie Bolton, Corneliu Buda, Nikolaos Diamantonis and Phil Davies, all of BP plc., for useful discussions. 

\section{Data availability statement}
The data that supports the findings of this study are available within the article.

\appendix \label{Appendix}
\section{Tables of results}
\subsection{Density at 25\degree C} 
	\begin{center}
		\begin{longtable}{|c|c|c|c|c|} % <-- Alignments: 1st column left, 2nd middle and 3rd right, with vertical lines in between
			\hline
			\textbf{Name}	&	\boldmath{$\rho_{\rm exp}$($\frac{\rm g}{\rm l}$)}	&	\boldmath{$\rho_{\rm model}$($\frac{\rm g}{\rm l}$)}	&	\boldmath{$\delta\rho_{\rm model}$($\frac{\rm g}{\rm l}$)}	&	\textbf{$\left|\Delta\right|$($\frac{\rm g}{\rm l}$)}	\\
			\hline									
			\hline									
			2,2-dimethylbutane	&	644.43	&	651.44	&	0.67	&	7.0	\\
			\hline									
			2,2-dimethylheptane	&	706.60	&	702.49	&	0.72	&	4.1	\\
			\hline									
			2,2-dimethylhexane	&	691.11	&	691.47	&	0.40	&	0.35	\\
			\hline									
			2,2-dimethyloctane	&	721.00	&	712.54	&	0.47	&	8.5	\\
			\hline									
			2,2-dimethylpentane	&	669.48	&	674.69	&	0.52	&	5.2	\\
			\hline									
			Decane	&	726.14	&	727.59	&	0.82	&	1.5	\\
			\hline									
			Heptane	&	679.50	&	678.40	&	0.42	&	1.1	\\
			\hline									
			Hexane	&	654.89	&	653.67	&	0.41	&	1.2	\\
			\hline									
			Nonane	&	713.75	&	714.01	&	0.97	&	0.26	\\
			\hline									
			Octane	&	698.76	&	699.37	&	0.49	&	0.61	\\
			\hline									
			3-ethylheptane	&	722.50	&	723.07	&	0.65	&	0.57	\\
			\hline									
			3-ethylhexane	&	709.45	&	709.08	&	0.77	&	0.37	\\
			\hline									
			3-ethyloctane	&	735.40	&	734.29	&	0.44	&	1.1	\\
			\hline									
			3-ethylpentane	&	693.92	&	690.89	&	0.52	&	3.0	\\
			\hline									
			3,3-diethylpentane	&	749.92	&	754.54	&	0.40	&	4.6	\\
			\hline									
			4-ethylheptane	&	722.30	&	722.38	&	0.37	&	0.08	\\
			\hline									
			4-ethyloctane	&	734.30	&	733.16	&	0.78	&	1.1	\\
			\hline									
			4-propylheptane	&	731.90	&	732.27	&	0.93	&	0.37	\\
			\hline									
			2-methylheptane	&	693.87	&	696.93	&	0.45	&	3.1	\\
			\hline									
			2-methylhexane	&	674.34	&	677.72	&	0.38	&	3.4	\\
			\hline									
			2-methylnonane	&	722.70	&	723.37	&	0.66	&	0.67	\\
			\hline									
			2-methyloctane	&	709.60	&	711.43	&	0.75	&	1.8	\\
			\hline									
			2-methylpentane	&	648.50	&	653.61	&	0.44	&	5.1	\\
			\hline									
			3-methylheptane	&	701.73	&	700.21	&	0.37	&	1.5	\\
			\hline									
			3-methylhexane	&	682.88	&	682.17	&	0.45	&	0.7	\\
			\hline									
			3-methylnonane	&	729.50	&	726.73	&	0.51	&	2.8	\\
			\hline									
			3-methyloctane	&	716.70	&	714.95	&	0.42	&	1.8	\\
			\hline									
			3-methylpentane	&	659.76	&	657.87	&	1.12	&	1.89	\\
			\hline									
			4-methylheptane	&	700.54	&	700.83	&	0.58	&	0.29	\\
			\hline									
			4-methylnonane	&	728.20	&	726.63	&	0.54	&	1.6	\\
			\hline									
			4-methyloctane	&	716.30	&	714.70	&	0.53	&	1.6	\\
			\hline									
			5-methylnonane	&	728.40	&	725.88	&	0.42	&	2.5	\\
			\hline									
			3-ethyl-2-methylheptane	&	739.80	&	729.11	&	0.43	&	11	\\
			\hline									
			3-ethyl-2-methylhexane	&	729.00	&	728.87	&	0.37	&	0.13	\\
			\hline									
			3-ethyl-2-methylpentane	&	715.20	&	717.26	&	0.51	&	2.1	\\
			\hline									
			3-ethyl-3-methylheptane	&	744.40	&	747.21	&	0.43	&	2.8	\\
			\hline									
			3-ethyl-3-methylhexane	&	736.00	&	736.93	&	0.53	&	0.93	\\
			\hline									
			3-ethyl-3-methylpentane	&	723.54	&	724.90	&	0.57	&	1.4	\\
			\hline									
			3-ethyl-4-methylheptane	&	746.60	&	742.77	&	0.31	&	3.8	\\
			\hline									
			3-ethyl-4-methylhexane	&	735.00	&	732.52	&	0.48	&	2.5	\\
			\hline									
			4-ethyl-2-methylheptane	&	732.20	&	734.77	&	0.37	&	2.6	\\
			\hline									
			4-ethyl-2-methylhexane	&	720.20	&	724.34	&	0.66	&	4.1	\\
			\hline									
			4-ethyl-3-methylheptane	&	746.80	&	742.22	&	0.33	&	4.6	\\
			\hline									
			4-ethyl-4-methylheptane	&	743.20	&	746.97	&	0.41	&	3.8	\\
			\hline									
			5-ethyl-2-methylheptane	&	731.50	&	733.73	&	0.40	&	2.2	\\
			\hline									
			5-ethyl-3-methylheptane	&	736.80	&	738.63	&	0.67	&	1.8	\\
			\hline									
			2,3-dimethylbutane	&	657.00	&	659.81	&	0.65	&	2.8	\\
			\hline									
			2,3-dimethylheptane	&	722.00	&	716.74	&	0.76	&	5.3	\\
			\hline									
			2,3-dimethylhexane	&	708.16	&	703.73	&	0.58	&	4.4	\\
			\hline									
			2,3-dimethyloctane	&	734.10	&	727.82	&	0.78	&	6.3	\\
			\hline									
			2,3-dimethylpentane	&	690.81	&	688.19	&	0.43	&	2.6	\\
			\hline									
			2,4-dimethylheptane	&	711.50	&	713.36	&	0.32	&	1.9	\\
			\hline								
			2,4-dimethylhexane	&	696.11	&	701.16	&	0.32	&	5.1	\\
			\hline									
			2,4-dimethyloctane	&	722.60	&	725.86	&	0.52	&	3.3	\\
			\hline									
			2,4-dimethylpentane	&	668.23	&	679.09	&	0.47	&	11	\\
			\hline									
			2,5-dimethylheptane	&	713.60	&	713.11	&	0.44	&	0.49	\\
			\hline									
			2,5-dimethylhexane	&	689.37	&	695.62	&	0.52	&	6.3	\\
			\hline									
			2,5-dimethyloctane	&	723.80	&	724.17	&	0.50	&	0.37	\\
			\hline									
			2,6-dimethylheptane	&	704.50	&	714.67	&	0.40	&	10.2	\\
			\hline									
			2,6-dimethyloctane	&	724.80	&	723.64	&	0.35	&	1.2	\\
			\hline									
			2,7-dimethyloctane	&	719.80	&	719.95	&	0.46	&	0.15	\\
			\hline									
			3,3-dimethylheptane	&	721.60	&	723.01	&	0.42	&	1.4	\\
			\hline									
			3,3-dimethylhexane	&	705.95	&	709.36	&	0.50	&	3.4	\\
			\hline									
			3,3-dimethyloctane	&	734.40	&	730.25	&	0.50	&	4.2	\\
			\hline									
			3,3-dimethylpentane	&	689.16	&	704.56	&	0.58	&	15	\\
			\hline									
			3,4-dimethylheptane	&	727.50	&	722.15	&	0.52	&	5.4	\\
			\hline									
			3,4-dimethylhexane	&	715.15	&	708.17	&	0.34	&	7.0	\\
			\hline									
			3,4-dimethyloctane	&	741.00	&	730.96	&	0.42	&	10	\\
			\hline									
			3,5-dimethylheptane	&	716.60	&	717.96	&	1.19	&	1.4	\\
			\hline									
			3,5-dimethyloctane	&	732.90	&	728.38	&	0.43	&	4.5	\\
			\hline									
			3,6-dimethyloctane	&	731.50	&	726.47	&	0.39	&	5.0	\\
			\hline									
			4,4-dimethylheptane	&	718.30	&	722.99	&	0.54	&	4.7	\\
			\hline									
			4,4-dimethyloctane	&	731.20	&	732.77	&	0.56	&	1.6	\\
			\hline									
			4,5-dimethyloctane	&	743.20	&	730.89	&	0.36	&	12	\\
			\hline									
		\caption{Results obtained from molecular dynamics simulations after correction factors in a leave-one-out cross validation are applied for density at 25\degree C. Experimental data is obtained from the TRC Thermodynamic tables \cite{TRC}.}
		\label{tab:density 25}
		\end{longtable} 	
	\end{center}

\subsection{Density at 100\degree C} 

	\begin{center}
		\begin{longtable}{|c|c|c|c|c|} % <-- Alignments: 1st column left, 2nd middle and 3rd right, with vertical lines in between
			\hline
			\textbf{Name}	&	\boldmath{$\rho_{\rm exp}$($\frac{\rm g}{\rm l}$)}	&	\boldmath{$\rho_{\rm model}$($\frac{\rm g}{\rm l}$)}	&	\boldmath{$\delta\rho_{\rm model}$($\frac{\rm g}{\rm l}$)}	&	\textbf{$\left|\Delta\right|$($\frac{\rm g}{\rm l}$)}	\\
			\hline									
			\hline									
			Decane	&	667.70	&	667.99	&	0.32	&	0.29	\\
			\hline									
			Heptane	&	611.00	&	612.63	&	0.47	&	1.6	\\
			\hline									
			Hexane	&	581.40	&	579.66	&	0.85	&	1.7	\\
			\hline									
			Nonane	&	652.50	&	652.14	&	0.67	&	0.36	\\
			\hline									
			Octane	&	635.19	&	635.37	&	0.60	&	0.18	\\
			\hline									
			2-methylheptane	&	632.00	&	631.69	&	0.52	&	1.7	\\
			\hline									
			2-methylhexane	&	602.00	&	611.99	&	0.63	&	8.8	\\
			\hline									
			2-methylpentane	&	574.30	&	580.04	&	0.70	&	4.5	\\
			\hline									
			3-methylheptane	&	638.40	&	636.39	&	0.66	&	3.4	\\
			\hline									
			3-methylhexane	&	619.00	&	614.50	&	0.65	&	6.0	\\
			\hline									
			3-methylpentane	&	582.40	&	586.35	&	0.52	&	2.7	\\
			\hline									
			4-methylheptane	&	639.00	&	635.54	&	0.53	&	4.9	\\
			\hline									
			3-ethylhexane	&	647.00	&	644.06	&	0.66	&	2.9	\\
			\hline									
			3-ethylpentane	&	621.00	&	625.14	&	0.60	&	4.1	\\
			\hline									
			4-propylheptane	&	673.40	&	672.20	&	0.44	&	1.2	\\
			\hline									
			2,3-dimethylbutane	&	582.50	&	584.21	&	0.74	&	1.7	\\
			\hline									
			2,3-dimethylhexane	&	644.10	&	635.12	&	0.57	&	9.0	\\
			\hline									
			2,3-dimethylpentane	&	626.00	&	627.51	&	0.62	&	1.5	\\
			\hline									
			2,4-dimethylhexane	&	616.30	&	632.27	&	0.52	&	16	\\
			\hline									
			2,4-dimethylpentane	&	601.00	&	605.48	&	0.52	&	4.5	\\
			\hline									
			2,5-dimethylhexane	&	623.60	&	625.31	&	0.48	&	1.7	\\
			\hline									
			2,6-dimethylheptane	&	640.00	&	644.74	&	0.42	&	4.7	\\
			\hline									
			2,7-dimethyloctane	&	660.20	&	656.44	&	0.86	&	3.8	\\
			\hline									
			3,3-dimethylhexane	&	646.70	&	640.78	&	1.05	&	5.9	\\
			\hline									
			3,3-dimethylpentane	&	608.00	&	635.21	&	0.45	&	27	\\
			\hline									
			3,4-dimethylhexane	&	658.50	&	639.86	&	0.64	&	19	\\
			\hline									
			4,5-dimethyloctane	&	685.50	&	665.46	&	0.63	&	20	\\
			\hline									
			2,2-dimethylbutane	&	568.30	&	576.37	&	0.75	&	8.1	\\
			\hline									
			2,2-dimethylhexane	&	626.10	&	618.23	&	0.83	&	7.9	\\
			\hline									
			2,2-dimethylpentane	&	601.90	&	601.71	&	0.46	&	0.19	\\
			\hline									
			3-ethyl-2-methylpentane	&	657.00	&	638.35	&	0.45	&	19	\\
			\hline									
			3-ethyl-3-methylhexane	&	641.00	&	661.82	&	0.43	&	21	\\
			\hline									
			3-ethyl-3-methylpentane	&	663.30	&	648.87	&	0.85	&	14	\\
			\hline									
			5-ethyl-2-methylheptane	&	672.70	&	658.79	&	0.60	&	14	\\
			\hline																
		
		\caption{Results obtained from molecular dynamics simulations after correction factors in a leave-one-out cross validation are applied for density at 100\degree C. Experimental data is obtained from the TRC Thermodynamic tables \cite{TRC}.}
		\label{tab:density 100}
		\end{longtable} 
		
	\end{center}

\subsection{Viscosity of linear alkanes}
\begin{table}[H]
	\begin{center}
		\begin{tabular}{|c|c|c|c|c|c|} % <-- Alignments: 1st column left, 2nd middle and 3rd right, with vertical lines in between
			\hline
			\textbf{Name} & \boldmath{$\rho_{exp}({\frac{g}{l}})$} &\boldmath{$\eta_{\rm exp}$} \textbf{(cSt)} & \boldmath{$\eta_{\rm pred}$} \textbf{(cSt)} & \boldmath{$\delta\eta_{\rm pred}$} \textbf{(cSt)}  & \textbf{$\Delta_{\%}$} \\ \hline
			Hexane & 659 &0.46 &0.43 & 0.04 &  -7\\ \hline
			Heptane &684 &0.60  &0.68 & 0.08 &  13\\ \hline
			Octane  &703 &0.78 &0.78 & 0.04 &   0 \\ \hline
			Nonane  &718 &0.99 &1.01 &  0.03 &   2\\ \hline
			Decane  &730 &1.24 &1.13 & 0.08 &   -9\\ \hline
			Undecane  &740 &1.60 &1.53 & 0.04 &   -4 \\ \hline
			Dodecane  & 749&2.00 &1.81 & 0.12 &   -9.5 \\ \hline
			Tridecane  &756  &2.38 &2.32 & 0.27 &  -2.5 \\ \hline
			Tetradecane & 762&3.01 &2.70 & 0.14 &   -10 \\ \hline
		\end{tabular} 
		\caption{Summary of viscosity simulations for linear alkanes at 20$^{\circ}$. Alkane's name, experimental value of kinematic viscosity, simulation result, its uncertainty, and percent error are presented.}
		\label{tab:Linear}
	\end{center}
\end{table}

\subsection{Viscosity of tridecane as a function of pressure at 60\degree C}

\begin{table}[H]
	\begin{center}
		\begin{tabular}{|c|c|c|c|c|c|} % <-- Alignments: 1st column left, 2nd middle and 3rd right, with vertical lines in between
			\hline
			\textbf{p(MPa)} & \boldmath{$\rho_{exp}({\frac{g}{l}})$} &\boldmath{$\eta_{\rm exp}$} \textbf{(cSt)} & \boldmath{$\eta_{\rm pred}$} \textbf{(cSt)} & \boldmath{$\delta\eta_{\rm pred}$} \textbf{(cSt)}  & \textbf{$\Delta_{\%}$} \\ \hline
			0.1 & 728 & 1.33 & 1.38 & 0.16  & 3.8 \\ \hline
			20  & 743 & 1.67 & 1.67 & 0.06  & 0\\ \hline
			40 & 757 & 2.02 & 2.09 &  0.18  & 3.5\\ \hline
			60 & 768 & 2.36 & 2.23 & 0.08  & -6 \\ \hline
			80 & 779 & 2.74 & 2.84 & 0.18  & 3.6\\ \hline
			100 & 788 & 3.12 & 3.37 & 0.16  & 8.0\\ \hline
		\end{tabular} 
		\caption{Results of viscosity simulations for tridecane at 60$^{\circ}$ as a function of pressure. Pressure, experimental value of kinematic viscosity, simulation result, its uncertainty, and percent error are presented.}
		\label{tab:TridecaneP}
	\end{center}
\end{table}

\subsection{Viscosity of octane, dodecane and octadecane as a function of temperature}

\begin{table}[H]
	\begin{center}
		\begin{tabular}{|c|c|c|c|c|c|} % <-- Alignments: 1st column left, 2nd middle and 3rd right, with vertical lines in between
			\hline
			\textbf{T($^{\circ}$C)} & \boldmath{$\rho_{exp}({\frac{g}{l}})$} & \boldmath{$\eta_{\rm exp}$} \textbf{(cSt)} & \boldmath{$\eta_{\rm pred}$} \textbf{(cSt)} & \boldmath{$\delta\eta_{\rm pred}$} \textbf{(cSt)}  & \textbf{$\Delta_{\%}$} \\ \hline
			-10  & 729 & 1.15 & 1.10 & 0.05  & -4\\ \hline
			0 & 721 & 0.99 & 1.01 & 0.02  & 2\\ \hline
			25 & 699 & 0.73 & 0.68 & 0.01  & -7 \\ \hline
			40  & 686 & 0.63 & 0.66 &  0.03  & 5\\ \hline
			60 & 669 & 0.53 & 0.50 & 0.05  & -6\\ \hline
			80  & 652 & 0.45 & 0.48 & 0.10  & 6.67 \\ \hline
			100  & 635 & 0.39 & 0.38 & 0.003  & -3\\ \hline
			125  & 618 & 0.33 & 0.34 & 0.02  & 3 \\ \hline
		\end{tabular} 
		\caption{Summary of viscosity simulations for octane as a function of temperature. Temperature, experimental value of kinematic viscosity, simulation result, its uncertainty, and percent error are presented.}
		\label{tab:OctaneT}
	\end{center}
\end{table} 

\begin{table}[H]
	\begin{center}
		\begin{tabular}{|c|c|c|c|c|c|} % <-- Alignments: 1st column left, 2nd middle and 3rd right, with vertical lines in between
			\hline
			\textbf{T($^{\circ}$C)} & \boldmath{$\rho_{exp}({\frac{g}{l}})$} & \boldmath{$\eta_{\rm exp}$} \textbf{(cSt)} & \boldmath{$\eta_{\rm pred}$} \textbf{(cSt)} & \boldmath{$\delta\eta_{\rm pred}$} \textbf{(cSt)}  & \textbf{$\Delta_{\%}$} \\ \hline
			40  & 734 & 1.46 & 1.57 & 0.06  & 8\\ \hline
			60  & 720 & 1.12 & 1.03 & 0.09  & -8\\ \hline
			80  & 704 & 0.90 & 0.87 & 0.04  & -3 \\ \hline
			100 & 690 & 0.73 & 0.79 &  0.11 & 8.2\\ \hline
			125 & 671 & 0.60 & 0.58 & 0.02  & -3\\ \hline
			150 & 651 & 0.50 & 0.54 & 0.03  & 8 \\ \hline
			175 & 630 & 0.42 & 0.47 & 0.03  & 12\\ \hline
			200 & 609 & 0.36 & 0.40 & 0.02  & 11 \\ \hline
		\end{tabular} 
		\caption{Summary of viscosity simulations for dodecane as a function of temperature. Temperature, experimental value of kinematic viscosity, simulation result, its uncertainty, and percent error are presented.}
		\label{tab:DodecaneT}
	\end{center}
\end{table}

\begin{table}[H]
	\begin{center}
		\begin{tabular}{|c|c|c|c|c|c|} % <-- Alignments: 1st column left, 2nd middle and 3rd right, with vertical lines in between
			\hline
			\textbf{T($^{\circ}$C)} & \boldmath{$\rho_{exp}({\frac{g}{l}})$} & \boldmath{$\eta_{\rm exp}$} \textbf{(cSt)} & \boldmath{$\eta_{\rm pred}$} \textbf{(cSt)} & \boldmath{$\delta\eta_{\rm pred}$} \textbf{(cSt)}  & \textbf{$\Delta_{\%}$} \\ \hline
			50 & 762 & 3.23 & 3.24 & 0.25  & 0.31\\ \hline
			75  & 744 & 2.14 & 2.02 & 0.18  & -5.7\\ \hline
			100 & 727 & 1.55 & 1.65 & 0.20  & 6.5 \\ \hline
			125 & 709 & 1.18 & 1.12 &  0.12  & -5.1\\ \hline
			150 & 691 & 0.93 & 0.96 & 0.05  & 3\\ \hline
			175 & 674 & 0.77 & 0.81 & 0.07  & 5 \\ \hline
			200 & 656 & 0.64 & 0.63 & 0.04  & -2\\ \hline
		\end{tabular} 
		\caption{Results of viscosity simulations for octadecane as a function of temperature. Temperature, experimental value of kinematic viscosity, simulation result, its uncertainty, and percent error are presented.}
		\label{tab:OctadecaneT}
	\end{center}
\end{table}

% If in two-column mode, this environment will change to single-column format so that long equations can be displayed. 
%\begin{widetext}
%$$\mbox{put long equation here}$$
%\end{widetext}

% Figures should be put into the text as floats. 
% Use the graphics or graphicx packages (distributed with LaTeX2e).
% Put the label that you will use with \ref{} command in the braces of the \label{} command.
%
% \begin{figure}
% \end{figure}

% \begin{table}
% \end{tabular}
% \end{table}

%\begin{acknowledgments}
% Put your acknowledgments here.
%\end{acknowledgments}

\bibliography{Draft.bbl}

\begin{thebibliography}{49}

\bibitem{21}
Allen, William and Rowley, Richard L.,
"Predicting the viscosity of alkanes using nonequilibrium molecular dynamics: Evaluation of intermolecular potential models,"
The Journal of Chemical Physics, 106(24):10273-10281 (1997).

\bibitem{TRC}
American Petroleum Institute. Research~Project 44 and Texas Engineering Experiment Station. Thermodynamics Research Center.
"TRC Thermodynamic Tables: Hydrocarbons,"
Thermodynamics Research Center, Texas Engineering Experiment Station Texas A \& M University System, (1986).

\bibitem{Caudwell2004}
Caudwell, D.R., Trusler, J.P.M., Vesovic, V., and  Wakeham, W.A.,
"The viscosity and density of n-dodecane and n-octadecane at pressures up to 200 {MPa} and temperatures up to 473 {K},"
International Journal of Thermophysics, 25(5):1339--1352 (2004).

\bibitem{doi:10.1021/je800417q}
Caudwell, D.R., Trusler, J.P.M., Vesovic V., and Wakeham, W.A.,
"Viscosity and density of five hydrocarbon liquids at pressures up to 200 {MPa} and temperatures up to 473 {K}"
Journal of Chemical \& Engineering Data, 54(2):359--366 (2009).

\bibitem{15}
Cho, Soowon, Jeong, Sodham, Kim, Jun Mo, and Baig, Chunggi,
"Molecular dynamics for linear polymer melts in bulk and confined systems under shear flow,"
Scientific Reports, 7(1): 9004 (2017).

\bibitem{1}
Cui, S.T., Cummings, P.T., Cochran, H.D., Moore, J.D., and  Gupta S.A. Gupta,
"Nonequilibrium Molecular Dynamics Simulation of the Rheology of Linear and Branched Alkanes,"
International Journal of Thermophysics, 19(2):449-459 (1998).

\bibitem{6}
Cui, S. T., Cummings, P. T., and Cochran, H. D.,
"The calculation of the viscosity from the autocorrelation function using molecular and atomic stress tensors,"
Molecular Physics, 88(6):1657-1664 (1996).

\bibitem{4}
Cui, S.T., Gupta, S.A., Cummings, P.T., and Cochran, H.D.,
"Molecular dynamics simulations of the rheology of normal decane, hexadecane, and tetracosane,"
The Journal of Chemical Physics, 105(3):1214-1220 (1996).

\bibitem{9}
Daivis, Peter J. and Evans, Denis J.,
"Comparison of constant pressure and constant volume nonequilibrium simulations of sheared model decane,"
The Journal of Chemical Physics, 100(1): 541-547 (1994).

\bibitem{TridecanePressure}
Daugé P., Baylaucq, A., Canet X.,  and  Boned, C.,
"High pressure viscosity and density measurements of the binary mixture tridecane + 2,2,4,4,6,8,8-heptamethylnonane,"
High Pressure Research,18(1-6):291-296 (2000).

\bibitem{ViscosityComparison}
De la Porte, J.J., and Kossack, C.A.,
"A liquid phase viscosity–temperature model for long-chain n-alkanes up to C64H130 based on the Free Volume Theory"
Fuel, 135, 156-164 (2014).

\bibitem{SLLOD}
Edberg, Roger, Evans, Denis J., and Moriss, G.P.,
"Constrained molecular dynamics: Simulations of liquid alkanes with a new algorithm"
The Journal of Chemical Physics,84, 6933-6939 (1986).

\bibitem{SLLOD}
Evans, Denis J. and Morriss, G. P.,
"Nonlinear-response theory for steady planar Couette flow,"
Phys. Rev. A, 30, 1528-1530 (1984).

\bibitem{14}
Ewen, J.P., Heyes, D.M., and Dini, D.,
"Advances in nonequilibrium molecular dynamics simulations of lubricants and additives,"
Friction, 6(4):349--386 (2018).

\bibitem{Compr}
Felsing, W. A. Felsing and Watson, George M.,
"The Compressibility of Liquid n-Octane,"
Journal of the American Chemical Society, 64(8), 1822-1823 (1942).

\bibitem{13}
Flyvbjerg, H., and Petersen, H.G.,
"Error estimates on averages of correlated data," 
The Journal of Chemical Physics,91(1):461-466 (1989).

\bibitem{StatisticalLearning}
Friedman, Jerome, Hastie, Trevor, and Tibshirani, Robert,
"The Elements of Statistical Learning," 
2009

\bibitem{11}
Hess, Berk,
"Determining the shear viscosity of model liquids from molecular dynamics simulations,"
The Journal of Chemical Physics, 116(1):209-217 (2002).

\bibitem{ANNHoss}
Hosseini, Sayed Mostafa, Pierantozzi Mariano, and Moghadasi Jalil, 
"Viscosities of some fatty acid esters and biodiesel fuels from a rough hard-sphere-chain model and artificial neural network,"
Fuel, 235, 1083 - 1091 (2019).

\bibitem{28}
Janke, Wolfhard,
"Statistical Analysis of Simulations: Data Correlations and Error Estimation,"
2002

\bibitem{23}
Khare, Rajesh, De Pablo, Juan, and Yethiraj, Arun,
"Rheological, thermodynamic, and structural studies of linear and branched alkanes under shear,"
The Journal of Chemical Physics, 107(17):6956-6964 (1997).

\bibitem{Kioupis3}
Kioupis, Loukas I. and Maginn, Edward J.,
"Impact of Molecular Architecture on the High-Pressure Rheology of Hydrocarbon Fluids,"
The Journal of Physical Chemistry B, 104, 7774-7783 (2000).

\bibitem{Kioupis2}
Kioupis, Loukas I. and Maginn, Edward J.,
"Molecular Simulation of Poly-α-olefin Synthetic Lubricants:  Impact of Molecular Architecture on Performance Properties,"
The Journal of Physical Chemistry B, 103, 10781-10790 (1999).

\bibitem{Kioupis1}
Kioupis, Loukas I. and Maginn, Edward J.,
"Rheology, dynamics, and structure of hydrocarbon blends: a molecular dynamics study of n-hexane/n-hexadecane mixtures",
Chemical Engineering Journal, 74, 129-146 (1999).

\bibitem{25}
Kondratyuk N. D., Lankin, A. V., Norman, G. E., and Stegailov, V. V.,
"Relaxation and transport properties of liquid n-triacontane,"
Journal of Physics: Conference Series, 653, 012107 (2015).

\bibitem{Lees-Edwards}
Lees, A. W. Lees and Edwards, S. F.,
"The computer study of transport processes under extreme conditions,"
J. Phys. C: Solid State Phys. 5 1921 (1972).

\bibitem{Lev1}
Levenberg, Kenneth,
"A method for the solution of certain non-linear problems in least squares,"
Quart. Appl. Math. 2, 164-168 (1944).

\bibitem{22}
Liu, Pinzhi, Yu, Hualong, Ren Ning, Lockwood Frances E., and Wang, Jane Q.,
"Pressure--Viscosity Coefficient of Hydrocarbon Base Oil through Molecular Dynamics Simulations,"
Tribology Letters, 60(3), 34 (2015).

\bibitem{8}
Liu, Pinzhi, Lu, Jie, Yu, Hualong, Ren, Ning, Lockwood, Frances E., and Wang, Jane Q.,
"Lubricant shear thinning behavior correlated with variation of radius of gyration via molecular dynamics simulations,"
The Journal of Chemical Physics, 147(8):084904 (2017).

\bibitem{2}
Maginn, Edward J., Messerly, Richard A., Carlson, Daniel J., Roe, Daniel R., and Elliott, Richard J.,
"Best Practices for Computing Transport Properties 1. Self-Diffusivity and Viscosity from Equilibrium Molecular Dynamics [Article v1.0],"
Living Journal of Computational Molecular Science, University of Colorado Boulder, 1:6324-- (2018).

\bibitem{Lev2}
Marquardt, Donald W.,
"An Algorithm for Least-Squares Estimation of Nonlinear Parameters,"
Journal of the Society for Industrial and Applied Mathematics, 11(2), 431–441 (1963).

\bibitem{C100}
Moore, J.D, Cui, S.T, Cochran H.D, and Cummings, P.T.,
"A molecular dynamics study of a short-chain polyethylene melt.: I. Steady-state shear"
Journal of Non-Newtonian Fluid Mechanics, 93(1): 83-99, 2000

\bibitem{19}
Moriss, Gary P. and Evans, Denis J.,
"A constraint algorithm for the computer simulation of complex molecular liquids,"
Computer Physics Communications, 62(2), 267-278 (1991).

\bibitem{Mundy1}
Mundy, Christopher J., Siepmann Ilja J., Klein, Michael L.,
"Decane under shear: A molecular dynamics study using reversible NVT-SLLOD and NPT-SLLOD algorithms,"
The Journal of Chemical Physics,103,23,10192-10200 (1995).

\bibitem{Mundy3}
Mundy, Christopher J., Klein, Michael L., and Siepmann, Ilja J.,
"Determination of the Pressure−Viscosity Coefficient of Decane by Molecular Simulation,"
J. Phys Chem 100, 16779 (1996).

\bibitem{Mundy2}
Mundy, Christopher J., Balasubramanian S., Bagchi, Ken, Siepmann Ilja J., and Klein, Michael L.,
"Equilibrium and non-equilibrium simulation studies of fluid alkanes in bulk and at interfaces,"
Faraday Discuss. 104, 17 (1996).

\bibitem{7}
Nevins, D. Nevins  and Spere, F. J.,
"Accurate computation of shear viscosity from equilibrium molecular dynamics simulations,"
Molecular Simulation, 33(15):1261-1266 (2007).

\bibitem{Lawrence}
Novak, Lawrence T.,
"Predictive Corresponding-States Viscosity Model for the Entire Fluid Region: n-Alkanes,"
Industrial \& Engineering Chemistry Research, 52, 20, 6841-6847 (2013).

\bibitem{3}
Pan Guoai and McCabe Clare,
"Prediction of viscosity for molecular fluids at experimentally accessible shear rates using the transient time correlation function formalism,"
The Journal of Chemical Physics, 125(19):194527 (2006).

\bibitem{5}
Payal, Rajdeep Singh, Balasubramanian S., Rudra, Indranil, Tandon Kunj, Mahlke Ingo, Doyle David, and Cracknell, Roger,
"Shear viscosity of linear alkanes through molecular simulations: quantitative tests for n-decane and n-hexadecane,"
Molecular Simulation, 38(14-15):1234-1241 (2012).

\bibitem{LAMMPS}
Plimpton, S.,
"Fast Parallel Algorithms for Short-Range Molecular Dynamics"
Journal of Computational Physics, 117, 1-19, (1995).

\bibitem{HardSphere}
Riesco, Nicolas and Vesovic, Velisa,
"Extended hard-sphere model for predicting the viscosity of long-chain n-alkanes,"
Fluid Phase Equilibria, 425, 385-392 (2016).

\bibitem{me}
Santak, Pavao, and Conduit, Gareth,
"Predicting physical properties of alkanes with neural networks,"
Fluid Phase Equilibria, 501, 112259 (2019).

\bibitem{PCFF}
Sun, Huai, Mumby, Stephen J., Maple, Jon R., Hagler, Arnold T.,
"An ab Initio CFF93 All-Atom Force Field for Polycarbonates"
Journal of the American Chemical Society, 16(7): 2978-2987 (1994).

\bibitem{COMPASS}
Sun H.,
"COMPASS:  An ab Initio Force-Field Optimized for Condensed-Phase Applications:Overview with Details on Alkane and Benzene Compounds,"
The Journal of Physical Chemistry B, 102, 7338-7364 (1998).

\bibitem{ANNSuzuki}
Suzuki, Takahiro, Ebert, Ralf-Uwe, and Schüürmann, Gerrit,
"Application of Neural Networks to Modeling and Estimating Temperature-Dependent Liquid Viscosity of Organic Compounds,"
Journal of Chemical Information and Computer Sciences, 41, 3, 776-790 (2001).

\bibitem{18}
Yang,Y., Pakkanen T.A., and Rowley, R.L.,
"Nonequilibrium Molecular Dynamics Simulations of Shear Viscosity: Isoamyl Alcohol, n-Butyl Acetate, and Their Mixtures,"
International Journal of Thermophysics, 21(3): 703--717 (2000).

\bibitem{16}
Yang,Y., Pakkanen T.A., and Rowley, R.L.,
"NEMD Simulations of Viscosity and Viscosity Index for Lubricant-Size Model Molecules, International Journal of Thermophysics,"
23(6):1441--1454 (2002).

\bibitem{12}
Zhang, Haizhong, and Ely, James F.,
"AUA model NEMD and EMD simulations of the shear viscosity of alkane and alcohol systems,"
Fluid Phase Equilibria, 217(1):111-118 (2004).

\end{thebibliography}

\end{document}